\documentclass[]{pasj01}
\usepackage{bm}
\usepackage{lineno}

\Received{}
 
\begin{document} 

\title{ The History of The Milky Way: The Evolution of Star Formation, Cosmic Rays,
Metallicity, and Stellar Dynamics over Cosmic Time}

%
%

\author{Jiro \textsc{shimoda}$^*$\altaffilmark{1,2}}
\altaffiltext{1}{Institute for Cosmic Ray Research, The University of Tokyo, \\
5-1-5 Kashiwanoha, Kashiwa, Chiba 277-8582, Japan}
\email{jshimoda@icrr.u-tokyo.ac.jp}

\author{Shu-ichiro \textsc{Inutsuka}\altaffilmark{2}}
\altaffiltext{2}{Department of Physics, Graduate School of Science, Nagoya University, \\
Furo-cho, Chikusa-ku, Nagoya 464-8602, Japan}

\author{Masahiro \textsc{Nagashima}\altaffilmark{3}}
\altaffiltext{3}{
Faculty of Education, Bunkyo University, Koshigaya, Saitama 343-8511, Japan}





\maketitle

\begin{abstract}
We study the long-term evolution of the Milky Way (MW) over cosmic time by modeling
the star formation, cosmic rays, metallicity, stellar dynamics, outflows and inflows
of the galactic system to obtain various insights into the galactic evolution. The mass
accretion is modeled by the results of cosmological $N$-body simulations for the cold
dark matter. We find that the star formation rate is about half the mass accretion rate
of the disk, given the consistency between observed Galactic Diffuse X-ray Emissions (GDXEs)
and possible conditions driving the Galactic wind. Our model simultaneously reproduces
the quantities of star formation rate, cosmic rays, metals, and the rotation curve of the
current MW. The most important predictions of the model are that there is an unidentified
accretion flow with a possible number density of $\sim10^{-2}~{\rm cm^{-3}}$ and the part of
the GDXEs originates from a hot, diffuse plasma which is formed by consuming about 10~\% of
supernova explosion energy. The latter is the science case for future X-ray missions; {\it XRISM},
{\it Athena}, and so on. We also discuss further implications of our results for the planet
formation and observations of externalgalaxies in terms of the multimessenger astronomy.
\end{abstract}

\section{Introduction}
\label{sec:intro}
The evolution of galaxies over cosmic time is related to many astrophysical subjects such as
the formation of stars, the origin of cosmic rays (CRs) and radiation, and the evolution of
the environment for the life on the planets, and has been widely studied (e.g., \cite{kruit11,
putman12,naab17}). Nevertheless, our current understanding is far from sufficient. The long-term
star formation rate of the Milky Way (MW), which continues with an almost constant rate of several
$M_\odot~{\rm yr^{-1}}$ throughout the cosmic age (e.g., \cite{haywood16}), is a representative
puzzle; if the constant star formation resulted from the similar Galactic disk conditions during
the cosmic age of $14$~Gyr, the gaseous matter with a mass of $\sim10^9~M_\odot$ would have been
depleted within $\sim1$~Gyr. This puzzle can be translated by the cosmological context. Supposing
the cosmic density ratio of the baryon to dark matter (DM), $\sim0.1$ (\cite{planck20}), the total
baryon mass of the MW is estimated to be $\sim10^{11}~M_\odot$, while the total mass of stars is
$4\mathchar`-6\times10^{10}~M_\odot$ (\cite{blandh16}). Therefore, we need to find an explanation
why the half of the gaseous matter is converted into the stars, leaving $\sim1$~\% of the mass in the
galactic disk. In this paper, we construct the galactic evolution model to elucidate what can be the
essence of this fine-tuning mechanism.
\par
The galactic wind (outflow from the disk) is invoked to explain the observed metal absorption lines
in the circumgalactic medium (CGM, \cite{tumlinson17}). \citet{shimoda22a} studied the galactic wind
considering the effects of the radiative cooling and CR diffusion for the case of the MW. They found
that the current conditions of the MW allow the existence of the wind with a mass transfer rate of
several $M_\odot~{\rm yr}^{-1}$ which is comparable to the star formation rate. Thus, the galactic wind
is also important for the mass budget of the Galactic system. They also pointed out that the mass loss
due to the wind and star formation should be balanced with the disk mass accretion to explain the constant
star formation rate over cosmic time. In this paper, following their considerations, we dedicate to find
the possible gas accretion to reproduce the current conditions of the MW.
\par
Since we consider the CR-driven wind and the metal-polluted CGM, our model also predicts the amounts of CRs
and metals, simultaneously. The combination of predictions on the star formation, the CRs, the metals, and
the gaseous matter distribution can be useful for the broad astrophysical subjects such as the planet formation
(e.g., \cite{tsukamoto22a,tsukamoto22b}) and particle astrophysics in terms of the multimessenger astronomy
(e.g., \cite{murase22}). Therefore, the implications for various observations are important to test our model.
These are also discussed in this paper.
\par
This paper is organized as follows: In section~\ref{sec:model}, we review the important physical processes
through the construction of our model and present analytical estimates of the current star formation rate,
metallicity, and CR energy density. The numerical results of our model are presented in section~\ref{sec:results}.
We discuss the implications for observations and future prospects in section~\ref{sec:implication} and summarize
our results in section~\ref{sec:summary}. We will find that the origins of the outflow and inflow are controversially
uncertain, although they are really important for understanding galactic evolution.

\section{Model Description}
\label{sec:model}
%
\begin{figure}[htbp]
\begin{center}
\includegraphics[scale=0.35]{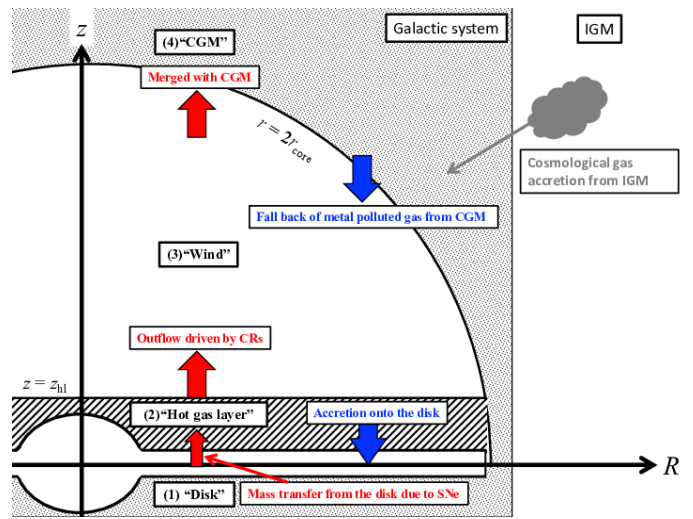}
\end{center}
\caption{Schematic illustration of our model. The horizontal axis shows the radial
distance $R$ in cylindrical coordinates. The vertical axis shows the vertical distance $z$.
The galactocentric radius is $r=\sqrt{R^2+z^2}$.}
\label{fig:illust}
\end{figure}
%
We model the galactic system under the assumption of axial symmetry. Figure~\ref{fig:illust} shows a schematic
diagram of our model. The horizontal axis shows the radial distance $R$ in cylindrical coordinates. The vertical
axis shows the vertical distance $z$. Thus, the galactocentric radius is $r=\sqrt{R^2+z^2}$. The model consists
of four parts: (1) the galactic disk, (2) the hot gas layer extending to $z=z_{\rm hl}\sim$~kpc which is responsible
for the observed diffuse X-ray emission~(e.g., \cite{nakashima18}), (3) the galactic wind extending to around the DM core
radius of $r=2~r_{\rm core}(t)$, and (4) the CGM extending from $r>2~r_{\rm core}(t)$. The red arrows indicate the outflow.
The SNe expel the diffuse gaseous matter from the disk into the hot gas layer. The CRs can drive the galactic wind
from $z=z_{\rm hl}=2$~kpc (\cite{shimoda22a}). We assume that the wind merges with the CGM at $r=2~r_{\rm core}(t)$. The
blue arrows indicate the inflow. We suppose that a fraction of the metal-polluted CGM condenses and falls to the disk
due to the radiative cooling.  The dynamics of the wind is estimated from the CR pressure gradient and the gravitational
acceleration. There is the cosmological gas accretion from the intergalactic medium (IGM). We assume that a fraction of
the accretion gas is expelled by the wind to remain in the CGM, and the other fraction falls onto the disk. In the following
we describe basic equations of the model for each part respectively. The model parameters are obtained by following the
various observations and theoretical studies for the MW.

\subsection{the cosmological accretion}
\label{sec:igm}
\begin{figure}[htbp]
\begin{center}
\includegraphics[scale=0.65]{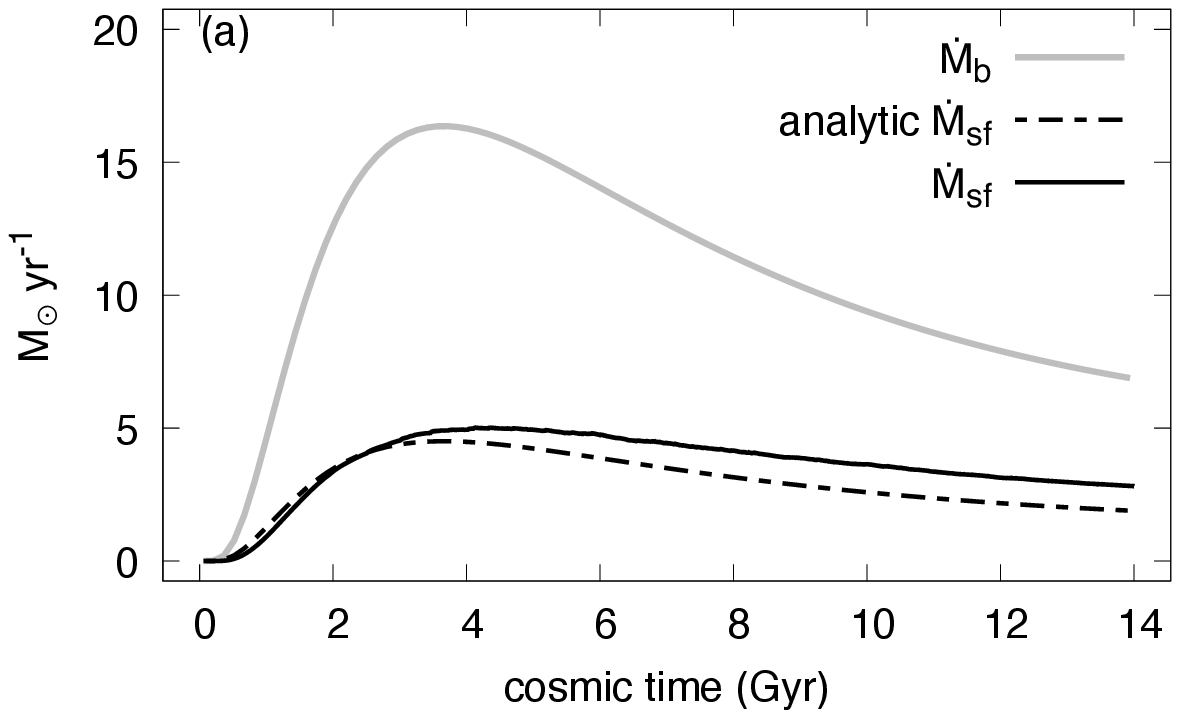}
\includegraphics[scale=0.65]{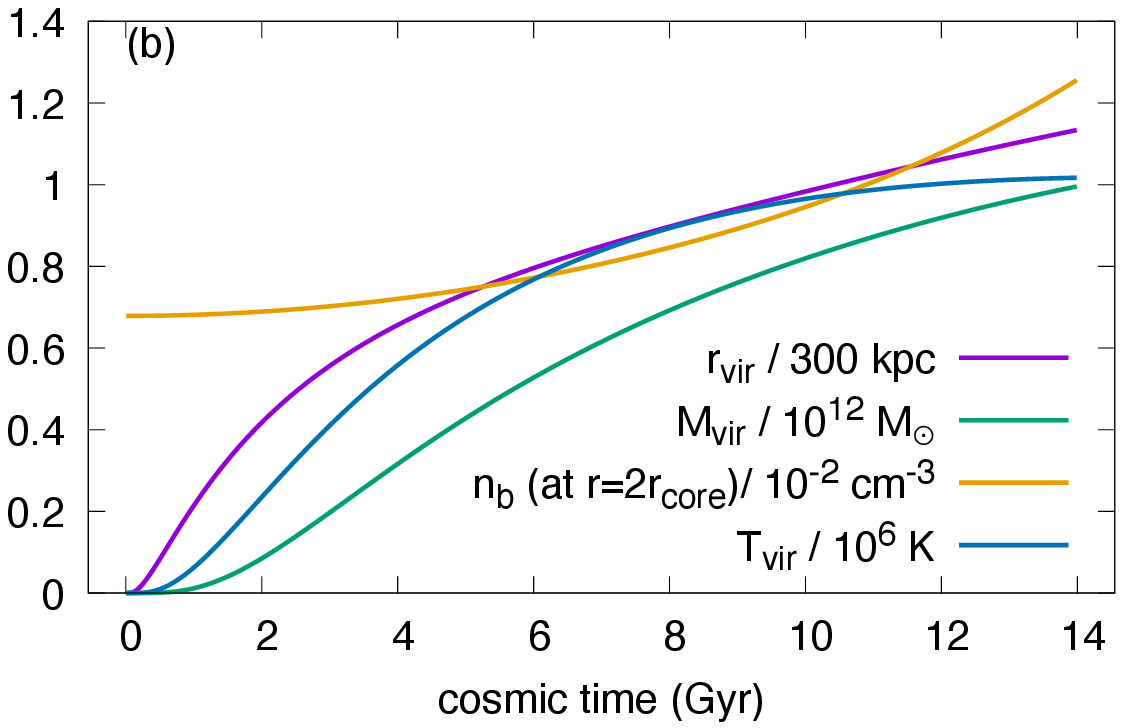}
\end{center}
\caption{(a) The solid gray line is the baryon accretion rate $\dot{M}_{\rm b}(t)$. The broken black line
is the analytical estimate of the star formation rate given by the equation (\ref{eq:analytic 2})
with setting $f_{\rm ms}=0.22$, $\eta_{\rm w}=6.58$, and $\dot{\Sigma}_{\rm cl}=\dot{\Sigma}_{\rm fb}=0$.
(see, section~\ref{sec:disk} and \ref{sec:wind}). The black solid line shows the numerical results of the
star formation rate for the fiducial model (see, section~\ref{sec:results}). (b) The purple line shows the
virial radius $r_{\rm vir}(t)$, which becomes $r_{\rm vir}\simeq342$~kpc at $t=14$~Gyr. The green line shows
the virial mass of the DM halo $M_{\rm vir}$. The number density of the accreting baryon at $r=2r_{\rm core}$
is shown by the orange line as $n_{\rm b}(t,2r_{\rm core})=f_{\rm cb}\rho_{\rm dm}(t,2r_{\rm core})/m_{\rm p}$.
The blue line shows the virial temperature which is given by the equation~(\ref{eq:T_vir}).}
\label{fig:Nbody}
\end{figure}

The galaxy is considered to be formed and evolved with the cosmological baryon accretion. In this paper, we presume that
the baryon accretion rate to the galactic system, $\dot{M}_{\rm b}(t)$, is proportional to the DM accretion rate, $\dot{M}_{\rm vir}$,
as $\dot{M}_{\rm b}=f_{\rm cb}\dot{M}_{\rm vir}$, where $f_{\rm cb}=0.156$ is the cosmic baryon fraction. We follow the results of
cosmological $N$-body simulations for the cold DM and their fitting functions given by \citet{rodriguez16} (the "Instantaneous" model
is used). The cosmological parameters are the same as in their settings. The fitting function of $\dot{M}_{\rm vir}$ is parametrized
by the current DM total mass $M_{\rm vir,0}$. We set $M_{\rm vir,0}=10^{12}~M_\odot$ (e.g., \cite{sofue12,posti19}).
Figures~\ref{fig:Nbody}a and \ref{fig:Nbody}b show $\dot{M}_{\rm b}$ and $M_{\rm vir}$, respectively.
%
%
\par
The mass density of the DM halo, $\rho_{\rm dm}(t,r)$, is assumed to be spherically symmetric with the NFW density
profile (\cite{nfw96,nfw97}). We assume the core radius to be $r_{\rm core}(t)=r_{\rm vir}(t)/20$, where $r_{\rm vir}(t)$
is the virial radius defined in the equation~(1) of \citet{rodriguez16}. In this paper, we assume that the mass density of the
accreting baryon is proportional to the DM mass density as $\rho_{\rm b}(t,r)=f_{\rm cb}\rho_{\rm dm}(t,r)$, which is shown in
Figure~\ref{fig:Nbody}b in terms of the number density $n_{\rm b}=\rho_{\rm b}/m_{\rm p}$ at the radius of $r=2r_{\rm core}$, where
$m_{\rm p}$ is the proton mass. The virial temperature, defined as
%
\begin{eqnarray}
\label{eq:T_vir}
k_{\rm B}T_{\rm vir} = \frac{ 2 }{ 3 }\frac{ G M_{\rm vir} m_{\rm p} }{ r_{\rm vir} },
\end{eqnarray}
%
is also shown, where $k_{\rm B}$ is the Boltzmann constant.

\subsection{the galactic disk and hot gas layer}
\label{sec:disk}
The galactic gas disk can be modeled by a viscous accretion disk. The local turbulent viscosity leads to the transport of
the angular momentum~(\cite{shakura73}). The disk wind and mass accretion are responsible for the change in total angular
momentum~(\cite{suzuki09,suzuki10,suzuki16}). We apply this disk model to the Galactic gas disk following \citet{suzuki16}. 
\par
Under the axisymmetric approximation, the equation of the continuity and the conservation of the angular momentum can be
written as~\citep{balbus98},
%
\begin{eqnarray}
\label{eq:continuous}
&& \frac{\partial\rho}{\partial t}
   +\frac{1}{R}\frac{\partial}{\partial R}\left(\rho R v_R\right)
   +           \frac{\partial}{\partial z}\left(\rho   v_z\right) = 0,
\end{eqnarray}
%
and
%
\begin{eqnarray}
\label{eq:ang momentum}
&& \frac{\partial}{\partial t}\left(\rho R v_\phi\right)
  +\frac{1}{R}\frac{\partial}{\partial R}\left[R^2\left(\rho v_R v_\phi - \frac{B_R B_\phi}{4\pi\rho}\right)\right]
\nonumber \\
&&
  +           \frac{\partial}{\partial z}\left[\rho R \left( v_\phi v_z - \frac{B_\phi B_z}{4\pi\rho}\right)\right] = 0,
\end{eqnarray}
%
respectively. $\rho$ is the mass density and the other symbols have their usual meaning. The azimuthal velocity is
decomposed into the mean galaxy rotation velocity ${\cal V}_\phi$ and the turbulent perturbation $\delta v_\phi$ as
$v_\phi= {\cal V}_\phi+\delta v_\phi$. We assume that the rotation velocity of the gaseous matter is given by the
equilibrium condition between the radial gravitational force, $g_{\rm R}=\partial \Phi/\partial R$, and centrifugal
force ${\cal V}_\phi{}^2/R$, where $\Phi$ is the gravitational potential and ${\cal V}_\phi =\sqrt{R\big|g_R\big|}$.
For the numerical calculation of $\Phi$, we consider the mass components of the disk gas, stars, and the DM halo.
The stars and the disk gas are approximated at the midplane ($z=0$). In this case, the potential can be calculated as
the sum of uniform rings of width $\Delta R$ at each discretized radius (see, \cite{krough82} and \cite{lass83}), where
$\Delta R$ is the interval of radial coordinates. Then, from the equation~(\ref{eq:ang momentum}) and multiples of the
equation~(\ref{eq:continuous}) and $R {\cal V}_\phi$, we obtain
%
\begin{eqnarray}
\label{eq:radial mass flux}
\rho R v_R \times \frac{1}{R} \frac{\partial}{\partial R}\left( R{\cal V}_\phi \right)
= - \left[ \frac{1}{R}\frac{\partial}{\partial R}\left(\rho R^2 \alpha_{R \phi} \right)
     +            \frac{\partial}{\partial z}\left(\rho R   \alpha_{\phi z} \right) \right]
\nonumber \\
\end{eqnarray}
%
where we approximate $\partial(\rho R v_\phi)/\partial t\approx R{\cal V}_\phi\partial\rho/
\partial t$ and use the $\alpha$
prescription (\cite{shakura73});
$\alpha_{R\phi} \equiv v_R \delta v_\phi - B_R B_\phi/4\pi\rho$ and
$\alpha_{R\phi} \equiv v_R \delta v_\phi - B_R B_\phi/4\pi\rho$.
When $\partial (R{\cal V}_\phi)/\partial R=0$ (there is no radial gradient in the specific
angular momentum), we regard $\partial \rho/\partial t=0$ from the equation~(\ref{eq:ang momentum}).
For $\partial (R{\cal V}_\phi)/\partial R\ne0$, substituting the equation~(\ref{eq:radial mass flux})
to the equation of the continuity, we obtain
%
\begin{eqnarray}
&& \frac{\partial\rho}{\partial t}
  -\frac{1}{R}\frac{\partial}{\partial R}
\left[
\frac{ R }{ {\cal V}_\phi + R {\cal V}_\phi{}' }
\left\{ \frac{1}{R}\frac{\partial}{\partial R}\left(\rho R^2 \alpha_{R \phi} \right)
     +            \frac{\partial}{\partial z}\left(\rho R   \alpha_{\phi z} \right) \right\}
\right]
\nonumber \\
&& +\frac{\partial}{\partial z}\left(\rho z\right) = 0,
\end{eqnarray}
%
where ${\cal V}_\phi{}'=\partial{\cal V}_\phi/\partial R$. Integrating this equation along the vertical direction $z$ from
the bottom surface ($z=-H/2$) to the top surface ($z=H/2$), we can derive
%
\begin{eqnarray}
&& \frac{\partial \left( R\Sigma \right) }{\partial t  }
  -\frac{1}{R}\frac{\partial}{\partial R}
\left[
\frac{ R }{ {\cal V}_\phi + R {\cal V}_\phi{}' }
\left\{ \frac{1}{R}\frac{\partial}{\partial R}\left(\Sigma R^2 C_s{}^2 \bar{\alpha}_{R \phi} \right)
   \right. \right. \nonumber \\
&& \left.  \left.
      +            \frac{\partial}{\partial z}\left(\Sigma R   C_s{}^2 \bar{\alpha}_{\phi z} \right) \right\}
\right]
= -\left[\rho R v_z \right]^{H/2}_{-H/2},
\end{eqnarray}
%
where $\Sigma=\int\rho dz$ is the surface mass density, $\bar{\alpha}_{R \phi}\equiv
\int\rho\alpha_{R \phi} dz/C_s{}^2\Sigma$, and $\bar{\alpha}_{\phi z}\equiv \rho R
\alpha_{\phi z}/C_s{}^2\Sigma$. $C_s$ is the sound speed of the disk gas. In this paper,
we approximate $\Sigma=\rho H$ and fix the disk thickness as $H=300$~{\rm pc} and $C_s=
10~{\rm km~s^{-1}}$ for simplicity. The $[\rho R v_z]$ term on the right-hand side represents
the effects of outflow and accretion. We denote these two effects separately as $-[\rho R v_z]=
-R\dot{\Sigma}_{\rm blown}+ R\dot{\Sigma}_{\rm acc}$. $\dot{\Sigma}_{\rm blown}$ indicates the
mass transfer rate from the disk to the hot gas layer due to the supernova explosions and
$\dot{\Sigma}_{\rm acc}$ indicates the net mass accretion rate. Adding the source terms due to
the star formation ($-\dot{\Sigma}_{\rm sf}$) and mass ejection by supernovae ($\dot{\Sigma_{\rm ej}}$),
we finally obtain the diffusion-convection type equation as
%
\begin{eqnarray}
\label{eq:disk gas}
&&
         \frac{\partial  }{\partial t  }\left( R\Sigma \right)
-{\cal D}_\Sigma\frac{\partial^2}{\partial R^2}\left( R\Sigma \right)
-{\cal V}_\Sigma\frac{\partial  }{\partial R  }\left( R\Sigma \right)
-{\cal Q}_\Sigma \left(R\Sigma\right)
\nonumber \\
&& = -R\dot{\Sigma}_{\rm blown} + R\dot{\Sigma}_{\rm acc} - R\dot{\Sigma}_{\rm sf} + R\dot{\Sigma}_{\rm ej}.
\end{eqnarray}
%
The coefficients ${\cal V}_\Sigma$, ${\cal D}_\Sigma$, and ${\cal Q}_\Sigma$ can be derived by
simple algebra. We assume that the $\alpha$-coefficients are constant; $\bar{\alpha}_{R\phi}=
\bar{\alpha}_{\phi z}=10^{-6}$. In the realistic Galactic disk, the mass transfer speed  may not
exceed the typical sound velocity $C_s\approx10~{\rm km~s^{-1}}$, which is a consequence of the
very efficient Ly$\alpha$ cooling of the neutral hydrogen atoms. The sound crossing time of the
galactic length scale is $\sim 10~{\rm kpc}/C_s\sim1$~Gyr,i.e. the radial mass transfer dose not
affect the evolution so much and the galactic radial profile depends strongly on the source terms.
Therefore, we choose the small $\alpha$-coefficients to safely reproduce the realistic situation.
\par
The mass consumption rate due to the star formation is estimated as $-\dot{\Sigma}_{\rm sf}=
-\epsilon_{\rm sf}\Sigma/\tau_{\rm sf}$, where $\tau_{\rm sf}$ and $\epsilon_{\rm sf}$ are the star
formation time and efficiency, respectively. From the state-of-the-art studies, the present day star
formation rate is determined by the molecular cloud formation that is driven through multiple compressions
of diffuse gas by supernovae (\cite{inoue09,inoue12,inutsuka15,hennebelle19,pineda22}). Thus, we set
the star formation time as $\tau_{\rm sf}=100~{\rm pc}/C_s\simeq9.8~{\rm Myr}$, where the length scale
of $100$~pc is assumed to be a typical size of local bubbles (e.g., \cite{zucker22} in the solar
neighborhood and \cite{watkins23} in NGC~628). By simply adopting the Kennicutt-Schmidt law~(\cite{kennicutt12}),
we set the star formation efficiency as
%
\begin{eqnarray}
\label{eq:ks}
\epsilon_{\rm sf} = 0.01 \left( \frac{ \Sigma }{ 10~M_\odot~{\rm pc^{-2}} } \right)^{0.4},
\end{eqnarray}
%
which results in $\dot{\Sigma}_{\rm sf}\propto\Sigma^{1.4}$ and the effective star formation
time scale as $\tau_{\rm sf}/\epsilon_{\rm sf}\simeq1~{\rm Gyr}(\Sigma/10~M_\odot~{\rm pc}^{-2})^{-0.4}$.
For the numerical calculation, we set a threshold mass density of $R\Sigma\Delta R=10^3~M_\odot$
below which $\dot{\Sigma}_{\rm sf}=0$.
\par
The surface mass density of long-lived low-mass stars and that of short-lived massive stars
are calculated separately as
%
\begin{eqnarray}
\label{eq:low mass star}
\frac{d\Sigma_*}{dt}=\left(1-f_{\rm ms}\right)\dot{\Sigma}_{\rm sf},
\end{eqnarray}
%
and
%
\begin{eqnarray}
\label{eq:ms}
\frac{d\Sigma_{\rm ms}}{dt}=f_{\rm ms}\dot{\Sigma}_{\rm sf}-\dot{\Sigma}_{\rm sn},
\end{eqnarray}
%
where $\Sigma_*$ and $\Sigma_{\rm ms}$ are the surface mass densities of the low-mass stars
and massive stars, respectively. We will discuss the treatments of the left-hand side terms,
$d\Sigma_*/dt$ and $d\Sigma_{\rm ms}/dt$, later. $f_{\rm ms}$ is a mass fraction of the massive
stars resulting in supernova explosions at the end of their life. $\dot{\Sigma}_{\rm sn}$ is the
supernova rate in terms of the surface mass density. The massive star fraction, $f_{\rm ms}$, is
estimated from the initial mass function by following the arguments of \citet{inutsuka15};
$f_{\rm ms} = \left(8~M_\odot / m_{*,{\rm min}} \right)^{2-\beta}$, where $\beta=2.35$ corresponds
to the Salpeter power-law index for a higher stellar mass part and $m_{*,{\rm min}}$ is the effective
minimum stellar mass parametrizing the shape of a lower stellar mass part. In this paper, we set
$m_{*,{\rm min}}=0.1~M_\odot$ and thus $f_{\rm ms}\simeq0.22$. Note that the number fraction becomes
$(8~M_\odot/m_{*,{\rm min}})^{1-\beta} \simeq2.6\times10^{-3}$. Then, we set $\dot{\Sigma}_{\rm sn}=
\Sigma_{\rm ms} / \tau_{*,{\rm ms}}$, where $\tau_{*,{\rm ms}}$ is a typical lifetime of the massive
stars. The lifetime is estimated from the mass-luminosity relation, $\tau_*=10~{\rm Gyr}(m_*/
1~M_\odot)^{-2.5}$, as
%
\begin{eqnarray}
\tau_{*,{\rm ms}}
&=&\left\{ \int_{8M_\odot}^\infty \frac{1}{\tau_*} d\left(\frac{m_*}{1~M_\odot}\right)\right\}^{-1}
\nonumber \\
&=& 35\times\left(\frac{8~M_\odot}{1~M_\odot}\right)^{-3.5}{\rm Gyr}
\nonumber \\
&\simeq& 24.2~{\rm Myr}.
\end{eqnarray}
%
Note that the event rate of the supernova can be estimated as $\dot{\Sigma}_{\rm sn}/\bar{m}_{*,{\rm ms}}
\approx f_{\rm ms}\dot{\Sigma}_{\rm sf}/\bar{m}_{*,{\rm ms}}$, where the average stellar mass of massive
stars (the ratio of total mass to total number) is $\bar{m}_{*,{\rm ms}}=(\beta-1)/(\beta-2) \times8~M_\odot
\simeq30.9~M_\odot$ and the steady-state approximation $d\Sigma_{\rm ms}/dt\approx0$ is used. Then, the total
event rate becomes $\dot{N}_{\rm sn}=f_{\rm ms}\dot{M}_{\rm sf}/ \bar{m}_{*,{\rm ms}}\simeq0.02~{\rm yr^{-1}}
(\dot{M}_{\rm sf}/3~M_\odot~{\rm yr^{-1}})(m_{*,{\rm min}}/0.1~M_\odot)^{\beta-2}$, where $\dot{M}_{\rm sf}=
2\pi\int\dot{\Sigma}_{\rm sf}R^2dR$ and $\beta=2.35$ is used. This is consistent with the current state of the MW.
\par
The supernova explosions result in the ejection of mass from $\Sigma_{\rm ms}$ to $\Sigma$, the rate of
which is given by $\dot{\Sigma}_{\rm ej}$ in the equation~(\ref{eq:disk gas}). We suppose that each of the
massive stars leaves one neutron star with a mass of $m_{*,{\rm ns}}=1.4~M_\odot$ as a result of the supernova.
The net mass ejected per one supernova event is $m_{\rm ej}=\bar{m}_{*,{\rm ms}}-m_{*,{\rm ns}}$ and the mass
ejection rate is $\dot{\Sigma}_{\rm ej}=(m_{\rm ej}/\bar{m}_{*,{\rm ms}})\dot{\Sigma}_{\rm sn}$. The surface
mass density of the created neutron stars is given by
%
\begin{eqnarray}
\label{eq:ns}
\frac{d\Sigma_{\rm ns}}{dt} = \frac{m_{*,{\rm ns}}}{\bar{m}_{*,{\rm ms}}}\dot{\Sigma}_{\rm sn}.
\end{eqnarray}
%
The treatment of the left-hand side term, $d\Sigma_{\rm ns}/dt$, will be discussed later.
\par
The supernovae may also blow the gaseous matter off the disk. We suppose that some of the blown gas is
responsible for the diffuse, X-ray emitting hot gas~(e.g., \cite{wada01,girichidis18}). The current
Galactic disk shows such diffuse X-ray emission, called the Galactic Diffuse X-ray Emissions (GDXEs).
Their origin is still under debated~(e.g., \cite{koyama18}, for reviews). Since the estimated temperature
of the GDXE ($>1$~keV) is much higher than the virial temperature of the current MW ($\sim0.1$~keV), we
expect that such X-ray emitting gas gose to the Galactic halo region. Indeed, the extended soft-X-ray emission
($\sim0.1$~keV) is observed at much higher latitudes than the case of GDXE (e.g., \cite{nakashima18,predehl20}).
Thus, we regard that the blown gas enters the hot gas layer which extends from $z=H/2$ to $z=z_{\rm hl}=2$~kpc.
The outflow rate from the disk into the layer, $\dot{\Sigma}_{\rm blown}$, in the right-hand side of the
equation~(\ref{eq:disk gas}) is estimated as
%
\begin{eqnarray}
\label{eq:sigma blown}
\dot{\Sigma}_{\rm blown}
&=&\eta_{\rm blown}
      \frac{ m_{\rm p}  }{ \bar{m}_{*,{\rm ms}} }
      \frac{ E_{\rm sn} }{ k_{\rm B}T_{\rm hl}          }
      \dot{\Sigma}_{\rm sn}
\equiv \eta_{\rm w}\dot{\Sigma}_{\rm sn},
\end{eqnarray}
%
where  $E_{\rm sn}=10^{51}~{\rm erg}$, and $T_{\rm hl}=3\times10^6$~K are the supernova explosion energy,
and the temperature of the hot gas layer, respectively. The conversion efficiency $\eta_{\rm blown}$ is
treated as a free parameter in this paper. The effective outflow efficiency driven by the supernovae,
$\eta_{\rm w}$, becomes
%
\begin{eqnarray}
\eta_{\rm w}
&\simeq& 6.58
\left( \frac{ \eta_{\rm blown}     }{ 0.1                 } \right)
\left( \frac{ E_{\rm sn}           }{ 10^{51}~{\rm erg}   } \right)
\nonumber \\
&\times&
\left( \frac{ T_{\rm hl}           }{ 3\times10^6~{\rm K} } \right)^{-1}
\left( \frac{ \bar{m}_{*,{\rm ms}} }{ 30.9~M_\odot        } \right)^{-1}.
\end{eqnarray}
%
Then, the surface mass density of the hot gas layer, $\Sigma_{\rm hl}$, is given by
%
\begin{eqnarray}
\frac{ d\Sigma_{\rm hl} }{ dt }
= \frac{ \dot{\Sigma}_{\rm blown} }{2} - \frac{ \Sigma_{\rm hl} }{ z_{\rm hl} } C_{s,{\rm hl}},
\end{eqnarray}
%
where $C_{s,{\rm hl}} =\sqrt{k_{\rm B}T_{\rm hl}/(0.6m_{\rm p})}\simeq200~{\rm km~s^{-1}}(T_{\rm hl}/
3\times10^6~{\rm K})^{1/2}$. The radial mass transfer in the hot gas layer is omitted for simplicity.
The factor $1/2$ of the first term on the right-hand side represents the existence of the layer on both
the $+z$ and $-z$ sides. The second term represents the mass loss due to the Galactic wind (discussed
later). The existence of the Galactic wind is supported by \citet{shimoda22a} for the current condition
of the MW.
\par
We can find the importance of $\eta_{\rm w}$ from the steady-state approximation for the
equation~(\ref{eq:disk gas}),
%
\begin{eqnarray}
 -R\dot{\Sigma}_{\rm blown} + R\dot{\Sigma}_{\rm acc} - R\dot{\Sigma}_{\rm sf} + R\dot{\Sigma}_{\rm ej}
\approx 0 \nonumber
\end{eqnarray}
%
and equation~(\ref{eq:ms}),
%
\begin{eqnarray}
\dot{\Sigma}_{\rm sf}f_{\rm ms}-\dot{\Sigma}_{\rm sn}
\approx 0. \nonumber
\end{eqnarray}
%
From these equations, we obtain $\dot{\Sigma}_{\rm ej}\approx (m_{\rm ej}/\bar{m}_{*,{\rm ms}})f_{\rm ms}\dot{\Sigma}_{\rm sf}$, $-
\dot{\Sigma}_{\rm blown}\approx-f_{\rm ms}\eta_{\rm w}\dot{\Sigma}_{\rm sf}$, and
%
\begin{eqnarray}
\label{eq:analytic 1}
\dot{\Sigma}_{\rm sf}
\approx \frac{ \dot{\Sigma}_{\rm acc} }{ 1 - f_{\rm ms} + f_{\rm ms}\eta_{\rm w} },
\end{eqnarray}
%
where we use $m_{\rm ej}/\bar{m}_{*,{\rm ms}}\simeq1$. The denominator is about $2.2$
for which $f_{\rm ms}\simeq0.22$ and $\eta_{\rm w} \simeq6.58$ ($f_{\rm ms}\eta_{\rm w}
\simeq1.42$). \citet{shimoda22a} showed that the mass transfer rate by the wind is comparable
to the current star formation rate of several $M_\odot~{\rm yr^{-1}}$, which is consistent with
the estimated $\dot{\Sigma}_{\rm blown}\simeq1.42\dot{\Sigma}_{\rm sf}$. Thus, our choice of
the conversion efficiency $\eta_{\rm blown}=0.1$ can be reasonable and the star formation rate
can be about half of the net accretion rate $\dot{\Sigma}_{\rm acc}$.  Note that the current
cosmological accretion rate of baryons estimated from the $N$-body simulations by \citet{rodriguez16}
is about $7~M_\odot~{\rm yr^{-1}}$ (see, Figure~\ref{fig:Nbody}a). Thus, the current star formation
rate can be $\sim 3~M_\odot~{\rm yr^{-1}}$.
Then, the total mass of the disk gas is $M\approx3\times10^9~M_\odot(\tau_{\rm sf}/10~{\rm Myr})
(\epsilon_{\rm sf}/0.01)^{-1}$ and the mass ratio of the hot gas layer to the disk gas is $\approx
\left( f_{\rm ms}\eta_{\rm w}z_{\rm hl}/C_{s,{\rm hl}} \right)\times3~M_\odot~{\rm yr^{-1}}/ M
\approx3\times10^{-3}(z_{\rm hl}/2~{\rm kpc})$ which is consistent with a typical number density of
the extended soft X-ray emission, $\sim 10^{-3}~{\rm cm^{-3}}$.
\par
We also calculate the metal surface mass density, $\Sigma_Z$, to estimate the metal pollution
of the CGM which is related to the net accretion rate $\dot{\Sigma}_{\rm acc}$ (discussed later).
The differential equation of the metal surface density is assumed to be the same form as the
equation~(${\ref{eq:disk gas}}$), except for $\dot{\Sigma}_{\rm ej}$ and $\dot{\Sigma}_{\rm blown}$.
The other differential equations of the stellar surface mass densities have the same form as those
of the total surface mass densities ($\Sigma_*\rightarrow\Sigma_{Z,*}$, $\Sigma_{\rm ms}
\rightarrow\Sigma_{Z,{\rm ms}}$, etc.). We include the newly created metal mass in the mass ejection
rate as
%
\begin{eqnarray}
\label{eq:metal ej}
\dot{\Sigma}_{Z, {\rm ej} }
= \left( \frac{ m_{\rm ej} }{ \bar{m}_{*,{\rm ms}} }
       + \frac{ m_{\rm co} }{ \bar{m}_{*,{\rm ms}} } \right)\dot{\Sigma}_{Z,{\rm sn}},
\end{eqnarray}
%
where $\dot\Sigma_{Z,{\rm sn}}=\Sigma_{Z,{\rm ms}}/\tau_{*,{\rm ms}}$. $m_{\rm co}$ is the CO core mass in a massive star. The mass ratio of the
massive star to the CO core is calculated as $\sim1/6\mathchar`-1/4$ (e.g., \cite{sukhold18,chieffi20}). In this paper, we fix $\bar{m}_{\rm co}/
\bar{m}_{*,{\rm ms}}=1/6$ for simplicity. The metal outflow rate is set to $\dot{\Sigma}_{Z, {\rm blown}} = (\Sigma_Z / \Sigma )
\dot{\Sigma}_{\rm blown}$.
\par
Here we describe the terms $d\Sigma_*/dt$, $d\Sigma_{\rm ms}/dt$, and $d\Sigma_{\rm ns}/dt$. For the short-lived massive-stars, we approximate
$d\Sigma_{\rm ms}/dt=\partial\Sigma_{\rm ms}/\partial t$. Since the gravitational potential evolves with time, the long-lived low-mass stars
and neutron stars can move away from their birthplace~(\cite{chandrasekhar43}). In this paper, we omit the motion along the $z$-direction for
simplicity. Then, for the numerical calculation, we introduce "parcels" consisting of the formed stars at each time step and each radial position
$(t^n,R_i)$ and solve for their motion;
%
\begin{eqnarray}
\frac{ d^2 R_{*,j} }{ dt^2 }
=-g_R(t,R_{*,j}) + \frac{ v_{*,\phi,j}{}^2 }{ R_{*,j} },
\end{eqnarray}
%
where $R_{*,j}$ and $v_{*,\phi,j}$ are the position and rotation velocity of the parcel denoted by the subscript $j$, respectively. Note that
$R_{*,j}v_{*,\phi,j}$ is constant. The net surface density of the mass is calculated as
%
\begin{eqnarray}
R\Sigma_*(t,R) = 
\sum_{j} \delta m_{*,j} \exp\left[ -\frac{ (R-R_{*,j})^2 }{ 2H^2 } \right],
\end{eqnarray}
%
where $\delta m_{*,j}$ is the line mass of parcel $j$. Let $t_{{\rm birth},j}$ be the birthtime of the stellar objects in the parcel $j$ at
$(t^n,R_i)$. Then, the birthplace is written as $R_{{\rm birth},j}=R_{*,j}(t_{{\rm birth},j})=R_i$. We assume the initial radial velocity as
$v_{*,R,j}(t_{{\rm birth},j})=dR_{*,j}/dt=0$ and the initial rotation velocity as $v_{*,\phi,j}(t_{\rm birth})={\cal V}_\phi(t_{\rm birth},
R_{\rm birth})$. The mass of parcel $j$ is given by $\delta m_{*,j}=R_{\rm birth}\dot{\Sigma}_{\rm sf}(t_{\rm birth},R_{\rm birth})\Delta t$
for the low-mass stars and $\delta m_{*,j}=R_{\rm birth}\dot{\Sigma}_{\rm ns}(t_{\rm birth},R_{\rm birth})\Delta t$ for the neutron stars,
respectively, where $\Delta t$ is the time interval of the numerical calculation. To save the computation costs, we implement that parcels $j$
and $k$ are unified under the conditions of $|R_{*,j}-R_{*,k}|<10^{-2}$, $|v_{*,\phi,j}-v_{*,\phi,k}|<10^{-2}$, $|v_{*,R,j}-v_{*,R,k}|<10^{-2}$,
and $v_{*,R,j}v_{*,R,k}>0$ to have a mass of $\delta m_{*,j}+\delta m_{*,k}$. The information of the birthtime and place are evaluated
as $(\delta m_{*,j}t_{{\rm birth},j}+\delta m_{*,k}t_{{\rm birth},k}) / (\delta m_{*,j}+\delta m_{*,k})$ and
   $(\delta m_{*,j}R_{{\rm birth},j}+\delta m_{*,k}R_{{\rm birth},k}) / (\delta m_{*,j}+\delta m_{*,k})$, respectively.
We set the inner and outer boundary conditions. For $R_{*,j}<\Delta R$, we regard the parcel to drop the galactic center $R=0$ and its motion
is no longer calculated. For $R_{*,j}>3R_{\rm B}$, where $R_{\rm B}=30$~kpc is the outer boundary of the gas disk calculation, we regard the
parcel to escape from the system.
\par
We summarize the disk model;
%
\begin{eqnarray}
\label{eq:summary disk}
\nonumber
&&         \frac{\partial  }{\partial t  }\left( R\Sigma \right)
  -{\cal D}_\Sigma\frac{\partial^2}{\partial R^2}\left( R\Sigma \right)
  -{\cal V}_\Sigma\frac{\partial  }{\partial R  }\left( R\Sigma \right)
\nonumber \\
&&
= -R\dot{\Sigma}_{\rm blown} + R\dot{\Sigma}_{\rm acc} - R\dot{\Sigma}_{\rm sf} + R\dot{\Sigma}_{\rm ej},
\nonumber \\
\nonumber
&& \frac{d\Sigma_*}{dt}=\left(1-f_{\rm ms}\right)\dot{\Sigma}_{\rm sf}, \\
\nonumber
&& \frac{d\Sigma_{\rm ms}}{dt}=f_{\rm ms}\dot{\Sigma}_{\rm sf}-\dot{\Sigma}_{\rm sn}, \\
\nonumber
&& \frac{d\Sigma_{\rm ns}}{dt} = \frac{m_{*,{\rm ns}}}{\bar{m}_{*,{\rm ms}}}\dot{\Sigma}_{\rm sn}, \\
\nonumber
&& \frac{ d\Sigma_{\rm hl} }{ dt }
= \frac{ \dot{\Sigma}_{\rm blown} }{2} - \frac{ \Sigma_{\rm hl} }{ z_{\rm hl} } C_{s,{\rm hl}}, \\
\nonumber
&& \dot{\Sigma}_{\rm sf} = \frac{ \epsilon_{\rm sf} }{ \tau_{\rm sf} }\Sigma, \\
\nonumber
&& \dot{\Sigma}_{\rm ej} = \frac{ m_{\rm ej} }{ m_{*,{\rm ms}} }\dot{\Sigma}_{\rm sn}
~~~ \left[ \dot{\Sigma}_{Z,{\rm ej}}
                         = \left( \frac{ m_{\rm ej} }{ \bar{m}_{*,{\rm ms}} }
                                + \frac{ m_{\rm co} }{ \bar{m}_{*,{\rm ms}} } \right)\dot{\Sigma}_{Z,{\rm sn}}
~({\rm for~metals}) \right], \\
\nonumber
&& \dot{\Sigma}_{\rm sn} = \frac{ \Sigma_{\rm ms} }{ \tau_{*,{\rm ms}} }, \\
&& \dot{\Sigma}_{\rm blown} = \eta_{\rm w}\dot{\Sigma}_{\rm sn}
~~~ \left[ \dot{\Sigma}_{Z,{\rm blown}} = \frac{ \Sigma_{Z} }{\Sigma}\dot{\Sigma}_{\rm blown}
~({\rm for~metals}) \right]. 
\end{eqnarray}
%
The metal surface mass densities are denoted by replacing the symbols as $\Sigma\rightarrow\Sigma_Z$, $\Sigma_*\rightarrow\Sigma_{Z,*}$,
$\dot{\Sigma}_{\rm sf}\rightarrow\dot{\Sigma}_{Z,{\rm sf}}=(\epsilon_{\rm sf}/\tau_{\rm sf})\Sigma_Z$ and so on, except for
$\dot{\Sigma}_{Z,{\rm ej}}$ and $\dot\Sigma_{Z,{\rm blown}}$. We find that $\Sigma$ is conserved (take $\dot{\Sigma}_{\rm blown}=
\dot{\Sigma}_{\rm scc}=0$ and $\Sigma_{\rm hl}=0$), while $\Sigma_Z$ increases at a rate of $(m_{\rm co}/\bar{m}_{*,{\rm ms}})
\dot{\Sigma}_{\rm sn}$.
\par
The estimation of the total mass of metal ejected by the supernovae clarifies
the importance of the galactic wind (i.e. the role of the CRs).
When the star formation continues with a constant rate of $\sim3~M_\odot~{\rm yr}^{-1}$ throughout the cosmic age,
the total metal mass
ejected by supernovae
becomes $\sim(m_{\rm co}/\bar{m}_{*,{\rm ms}})f_{\rm ms}\times3~M_\odot~{\rm yr}^{-1}\times14~{\rm Gyr}
\sim1.54\times10^9~M_\odot$, where $\dot{\Sigma}_{\rm sn}\approx f_{\rm ms}\dot{\Sigma}_{\rm sf}$ is used.
The total mass of the disk gas, $\sim3\times10^9~M_\odot$, which is comparable to the estimated metal mass, implies the existence of the wind;
the almost all metals should be expelled from the disk to simultaneously explain the total mass of the disk gas and the metallicity of $Z_\odot
\sim0.01$ in the current Galactic disk.
If the metal polluted gas is well mixed with the primordial gas with a mass of
$f_{\rm cb}M_{\rm vir,0}\sim10^{11}~M_\odot$, the mean metallicity becomes $\sim10^9~M_\odot/10^{11}_\odot\sim0.01$,
which is consistent with the solar metallicity. Thus, the CRs also play a key role in controling an amount of the metals
in the Galactic disk by driving the wind.

\subsection{the galactic halo region; wind, CGM, and IGM}
\label{sec:wind}
We suppose that the galactic wind is driven by CRs from the upper boundary of the hot gas layer, $z=z_{\rm hl}=2$~kpc, with a rate of
$( \Sigma_{\rm hl} / z_{\rm hl} ) C_{s,{\rm hl}}$ following \citet{shimoda22a}. The CRs are assumed to be accelerated by supernova remnant
shocks. In the following, we discuss the CR energy density and describe the numerical modeling of the galactic halo gas. In our model, the
disk accretion rate, $R\dot{\Sigma}_{\rm acc}$, is decomposed into three components as
%
\begin{eqnarray}
R\dot{\Sigma}_{\rm acc} = R\dot{\Sigma}_{\rm b,disk} + R\dot{\Sigma}_{\rm fb}+R\dot{\Sigma}_{\rm cl},
\end{eqnarray}
%
where the $\dot{\Sigma}_{\rm b,disk}$ is the cosmological baryon accretion rate on the disk, $\dot{\Sigma}_{\rm fb}$ is the mass coming from
the wind region, and $\dot{\Sigma}_{\rm cl}$ is the mass accretion rate from the metal-polluted CGM, respectively. Note that our model calculation
of the disk is done by the line mass of $R\Sigma$ as shown in the equation~(\ref{eq:summary disk}).
\par
\citet{shimoda22b} recently suggested that the energy injection rate of CRs at the shocks can be about 10~\% of the supernova explosion energy by
modeling the ion heating process at the shock transition. We follow their results and set the CR energy injection rate (${\rm erg~s^{-1}~cm^{-3}}$)
as
%
\begin{eqnarray}
\label{eq:cr injection}
\dot{q}_{\rm cr} = \frac{ \eta_{\rm cr} E_{\rm sn} }{ H \bar{m}_{*,{\rm ms}} } \dot{\Sigma}_{\rm sn},
\end{eqnarray}
%
where $\eta_{\rm cr}\sim0.1$ is the injection efficiency. The energy loss rate of the CRs via the hadronic interactions (${\rm erg~s^{-1}~cm^{-3}}$)
can be approximated by
%
\begin{eqnarray}
\label{eq:cr loss}
-{\cal L}_{\rm cr}= - \Lambda_{\rm h}\frac{ \Sigma }{ m_{\rm p}H } e_{\rm cr},
\end{eqnarray}
%
where $\Lambda_{\rm h}=7.44\times10^{-16}~{\rm s^{-1}~cm^3}$ is the collision rate per particle (see, \cite{pfrommer17} for details), and
$e_{\rm cr}~({\rm erg~cm^{-3}})$ is the energy density of the CRs. We presume that the acceleration time of CRs at each supernova remnant
shock is quite short, at most $\sim10^4$~yr, which is a typical radiative cooling time of the shock-heated plasma for an ambient density of
$\sim1~{\rm cm^{-3}}$ (e.g., \cite{vink06}). Thus, we approximate the energy density of the CRs around their source as $\dot{q}_{\rm cr}-
{\cal L}_{\rm cr}\approx0$, i.e.,
%
\begin{eqnarray}
\label{eq:e_cr}
e_{\rm cr,source} = \frac{ \eta_{\rm cr} E_{\rm sn} }{ \Lambda_{\rm h} }
\frac{ m_{\rm p}       }{ \bar{m}_{*,{\rm ms}} }
\frac{ \dot{\Sigma}_{\rm sn} }{ \Sigma               }.
\end{eqnarray}
%
Note that $e_{\rm cr,source}$ is assumed to be uniform along the vertical direction $z$ in the disk. The injected CRs will escape from the
galaxy. In this paper, the escape flux is assumed to be dominated by the diffusion effect (the convection effect is omitted for simplicity).
Then, we approximate the CR energy density released around the sources as
%
\begin{eqnarray}
\label{eq:e_cr out}
{\cal D}_{\rm cr}{\bm \nabla}^2 e_{\rm cr} = \frac{ e_{\rm cr,source} }{ \tau_{\rm cr,esc} },
\end{eqnarray}
%
where $\cal{D}_{\rm cr}$ and $\tau_{\rm cr,esc}$ are the CR diffusion coefficient and the CR escape time from the galaxy, respectively. The
escape time is estimated by the abundance of CR nuclei (e.g., the Boron-to-Carbon ratio) as $\sim10$~Myr. The diffusion coefficient is estimated
from the abundance ratio with the assumption that the CR scale height of $\sim$~kpc; ${\cal D}_{\rm cr}=3\times10^{28}~{\rm cm^2~s^{-1}}$ is
frequently used~(e.g., \cite{gabici19}). In this paper, we parametrize the CR scale height, $H_{\rm cr}=\sqrt{{\cal D}_{\rm cr}\tau_{\rm cr,esc}}$.
Then, the CR energy densities are estimated by using the steady-state approximation of $\dot{\Sigma}_{\rm sn}\approx f_{\rm ms}\dot{\Sigma}_{\rm sf}$
and $\Sigma\approx(\tau_{\rm sf}/\epsilon_{\rm sf})\dot{\Sigma}_{\rm sf}$ as
%
\begin{eqnarray}
e_{\rm cr,source}
&\approx&
\frac{ \eta_{\rm cr}E_{\rm sn} }{ \tau_{\rm sf}\Lambda_{\rm h} }
\frac{ m_{\rm p}               }{ \bar{m}_{*,{\rm ms}}         } f_{\rm ms}\epsilon_{\rm sf}
\nonumber \\
&\approx&
16~{\rm eV~cm^{-3}}
\left( \frac{     \eta_{\rm cr}               }{         0.1  } \right)
\left( \frac{        f_{\rm ms}               }{         0.22 } \right)
\nonumber \\
&\times&
\left( \frac{ \tau_{\rm sf}/\epsilon_{\rm sf} }{ 1~{\rm Gyr}  } \right)^{-1}
\left( \frac{ \bar{m}_{*,{\rm ms}}            }{ 30.9~M_\odot } \right)^{-1},
\label{eq:cr source}
\end{eqnarray}
%
and by using $\bm{\nabla}^2e_{\rm cr}\approx e_{\rm cr}/({\rm kpc})^2$ as,
%
\begin{eqnarray}
\label{eq:cr analytic}
e_{\rm cr}\approx 1.8~{\rm eV~cm^{-3}}
\left( \frac{ e_{\rm cr,source} }{ 16~{\rm eV~cm^{-3}} } \right)
\left( \frac{ H_{\rm cr}        }{  3~{\rm kpc}        } \right)^{-2}.
\end{eqnarray}
%
Thus, we can find that the CR energy density in the interstellar medium (ISM), which is expected to be a few ${\rm eV~cm^{-3}}$, is given
by $\eta_{\rm cr}/ {\cal D}_{\rm cr}\tau_{\rm esc}=\eta_{\rm cr}/H_{\rm cr}{}^2$. We treat the CR scale height $H_{\rm cr}$ as a free
parameter to study the effects of CR and fix the injection efficiency as $\eta_{\rm cr}=0.1$. In the CR injection scenario proposed by
\citet{shimoda22b}, the injection efficiency depends on the sonic Mach number of the collisionless shocks. Therefore, we regard that the
efficiency $\eta_{\rm cr}$ is universally determined by the local physics. On the other hand, the diffusion coefficient may depend on the
local magnetic field strength whose evolution over cosmic time may still be uncertain. Moreover, there is no consensus on the general
expression of the CR diffusion coefficient. Note that the energy densities depend on the star formation rate via $(\tau_{\rm sf}/
\epsilon_{\rm sf})^{-1}\propto\Sigma^{0.4}$; a larger $\Sigma$ results in a larger star formation rate and a larger CR energy density.
To estimate the CR energy density, we omit non-trivial numerical factors, so for simplicity we chose $H_{\rm cr}=3$~kpc. From the
numerical results discussed later, we find that $(\eta_{\rm cr},H_{\rm cr})=(0.1,10~{\rm kpc})$ produces a reasonable result. The height
of $H_{\rm cr}=10$~kpc is consistent with recent theoretical model of CR transport~(\cite{evoli20}).
\par
The galactic wind is driven by the CR pressure, $P_{\rm cr}=(\gamma_c-1)e_{\rm cr}$, where $\gamma_c=4/3$, from the hot gas layer.
\citet{shimoda22a} showed that for the current conditions of the MW, the wind density and temperature can be nearly constant at
$z\lesssim10$~kpc due to the balance between the radiative cooling and effects of CR heating. The thermal gas pressure and the CR
pressure are comparable to each other. From these results, we simplify the wind dynamics as
%
\begin{eqnarray}
\label{eq:wind}
\frac{d\bm{v}_{\rm w}}{dt}
=-\frac{\bm{\nabla}P_{\rm cr}}{\rho_{\rm w}} - \bm{g},
\end{eqnarray}
%
where we approximate the total pressure as $P\approx P_{\rm cr}$. $\rho_{\rm w}$ and $\bm{v}_{\rm w}$ are the mass density and velocity
of the wind, respectively. $\bm{g}$ is the gravitational acceleration calculated from the Poisson equation with the DM component and the
disk mass components ($\Sigma$, $\Sigma_*$, $\Sigma_{\rm ms}$, and $\Sigma_{\rm ns}$). For numerical calculations of the wind, we treat
the fluid parcel as a test particle denoted by the subscript $k$ and adopt the equation~(\ref{eq:wind}) as its equation of motion under
the axisymmetric approximation. The density $\rho_{{\rm w},k}$ is assumed to be constant. For each time step, new particles are newly
introduced at each point of $(R_i,z_{\rm hl})$. The initial conditions of the particles are assumed to be $\rho_{{\rm w},k}=\Sigma_{\rm hl}
(t,R)/z_{\rm hl}$, $v_{{\rm w}, R,k}=0$, $v_{{\rm w},\phi,k}=\sqrt{Rg_R(t,R,z_{\rm hl})}$, and $v_{{\rm w},z,k}=C_{\rm s,hl}$, where $g_R$
is the radial component of $\bm{g}$ and the centrifugal force equilibrium is assumed, $v_{{\rm w},\phi}{}^2/R=g_R$. The mass of each particle
is assumed to be $dm_{{\rm w},k}=\rho_{{\rm w},k}C_{s,{\rm hl}}R\Delta R\Delta t d\phi=$constant. The infinitesimal azimuthal element vanishes
by the axisymmetric integration. Using the particles located at $|z|<H/2$ and $R<R_{\rm B}$ with a density of $\rho_{{\rm w},k}\le\Sigma(t,R_k)
/H$, where $R_{\rm B}=30$~kpc is the outer boundary of the disk region, we calculate the accretion rate component, $R\dot{\Sigma}_{\rm fb}$, as
%
\begin{eqnarray}
\label{eq:fb}
R\dot{\Sigma}_{\rm fb}(t,R)
=
2\int_0^{2\pi} \sum_k \frac{dm_{{\rm w},k}}{d\phi}
\frac{N_k}{\Delta t}
\exp\left[ - \frac{ (R-R_k)^2 }{ 2H^2 } \right]
d\phi,
\nonumber \\
\end{eqnarray}
%
where $N_k=(\int_0^{R_{\rm B}} e^{- (R-R_k)^2 / 2H^2 } dR)^{-1}$ is the normalization factor. Note that the metal mass is calculated by the
same manner as the total mass. 
\par
We regard that the particles located at $r_k>r_{\rm cgm}$ enter the CGM region, where $r_k=\sqrt{R_k{}^2+z_k{}^2}$ and $r_{\rm cgm}=
2~r_{\rm core}(t)$ is the boundary galactocentric radius between the wind region and CGM region. The total CGM mass, $M_{\rm cgm}(t)$,
is calculated as
%
\begin{eqnarray}
\label{eq:cgm}
\frac{dM_{\rm cgm}}{dt}
= \dot{M}_{\rm w} + \dot{M}_{\rm ex}
- \dot{M}_{\rm cl},
\end{eqnarray}
%
where
%
\begin{eqnarray}
\dot{M}_{\rm w}= 2\int_0^{2\pi} \sum_k \frac{ dm_{{\rm w},k} }{ d\phi } d\phi,
\end{eqnarray}
%
and $\dot{M}_{\rm ex}$ represents the mass supply due to the interaction between the wind and the baryon accretion flow. Some of the
accreting baryons from the IGM may be expelled by the wind entering the CGM region. For simplicity, we assume that $\dot{M}_{\rm ex}=
\dot{M}_{\rm w}$. $\dot{M}_{\rm cl}$ is the mass loss of the CGM due to the metal pollution which results in a large radiative cooling
rate. For the CGM of externalgalaxies, absorption lines of lower ionized species such as H\emissiontype{I}, C\emissiontype{II}, and
Mg\emissiontype{II} are observed (\cite{tumlinson17}, and references therein). The existence of such lower ionized species implies the
existence of condensation phenomena that shield the cool gas from ionizing photons (e.g., the metagalactic radiation field, see \cite{gnat17}).
We assume that such cool, condensed gas leaves from the CGM by losing its angular momentum at a rate of 
%
\begin{eqnarray}
\label{eq:mdot_cl}
- \dot{M}_{\rm cl}
=-\frac{ \epsilon_{\rm cgm}             }{ \tau_{\rm cool}                      } M_{\rm cgm}
=-\frac{ \epsilon_{\rm cgm} kT_{\rm cgm}}{ n_{\rm cgm}\Lambda_{{\rm rad},\odot} }
  \frac{ Z_{\rm cgm} }{ Z_\odot }
  M_{\rm cgm},
\end{eqnarray}
%
where $n_{\rm cgm}$ is the mean number density estimated as
%
\begin{eqnarray}
\label{eq:n_cgm}
n_{\rm cgm} = \frac{3}{4\pi r_{\rm vir}{}^3}\frac{M_{\rm cgm}}{m_{\rm p}},
\end{eqnarray}
%
and the radiative cooling time is estimated as
%
\begin{eqnarray}
\label{eq:t_cool}
\tau_{\rm cool} = \frac{ n_{\rm cgm}\Lambda_{{\rm rad},\odot} }{ kT_{\rm cgm} }
\frac{  Z_\odot }{  Z_{\rm cgm}  }.
\end{eqnarray}
%
The mean temperature of the CGM, $T_{\rm cgm}$, is assumed to be equal to the virial temperature, $T_{\rm vir}(t)$, which is given by the
equation~(\ref{eq:T_vir}). $\Lambda_{{\rm rad},\odot}(T_{\rm cgm})$ is the radiative cooling rate for the solar abundance given by
\citet{shimoda22a}. $Z_{\rm cgm}$ is the metallicity of the CGM estimated as $Z_{\rm cgm}=M_{Z,{\rm cgm}}/M_{\rm cgm}$ and $Z_\odot=0.0134$
is the solar metallicity (\cite{asplund09}). $\epsilon_{\rm cgm}$ represents the efficiency of the angular momentum loss which is required
for both condensation and accretion phenomena. We fix the efficiency $\epsilon_{\rm cgm}=0.01$ by analogy with $\epsilon_{\rm sf}$. The factor
$Z_{\rm cgm}/Z_\odot$ approximately reflects the metallicity dependence of $\Lambda_{\rm rad}$.
\par
The net baryon accretion rate onto the disk is given by $\dot{M}_{\rm b}-\dot{M}_{\rm ex}+\dot{M}_{\rm cl}$.
The local accretion rate at $R$ is assumed to be
%
\begin{eqnarray}
\label{eq:sdot_b,disk}
R\dot{\Sigma}_{\rm b,disk}(t,R)
&=&
\frac{\dot{M}_{\rm b}-\dot{M}_{\rm ex}+\dot{M}_{\rm cl}}{4\pi}
\nonumber \\
&\times&
N_{\rm b,disk}
\exp\left[
-\frac{ (R-r_{\rm core})^2 }{ 2 r_{\rm core}{}^2 }
\right],
\end{eqnarray}
%
where $N_{\rm b,disk}=(\int_0^{R_{\rm B}} e^{-(R-r_{\rm core})^2/2r_{\rm core}{}^2}dR)^{-1}$ is the normalization factor. Here, the core radius
$r_{\rm core}$ is assumed to reflect a small angular momentum of the accreting gas from the IGM so that the galactic disk is formed. The radial
profile of the galactic disk depends strongly on the source terms as we discussed in the equation~(\ref{eq:disk gas}). Since the galactic wind
acts after the mass supply at a local point, the radial profile is almost given by this assumed $R\dot{\Sigma}_{\rm b.disk}$.
\par
The estimate of the star formation rate under the steady-state approximation discussed in the equation~(\ref{eq:analytic 1}) is rewritten as
%
\begin{eqnarray}
\label{eq:analytic 2}
\dot{\Sigma}_{\rm sf}
\approx \frac{ \dot{\Sigma}_{\rm b}+\dot{\Sigma}_{\rm fb}+\dot{\Sigma}_{\rm cl} }
{ 1 - f_{\rm ms} + 2f_{\rm ms}\eta_{\rm w} },
\end{eqnarray}
%
where we use the total disk mass accretion rate $\dot{\Sigma}_{\rm acc}=\dot{\Sigma}_{\rm b}-\dot{\Sigma}_{\rm w}+\dot{\Sigma}_{\rm cl}+
\dot{\Sigma}_{\rm fb}$ and approximate that $\dot{\Sigma}_{\rm w}\approx\dot{\Sigma}_{\rm blown}$. The dashed line in Figure~\ref{fig:Nbody}a
shows the estimated total star formation rate with $f_{\rm ms}=0.22$, $\eta_{\rm w}=6.58$, and $\dot{\Sigma}_{\rm cl}=\dot{\Sigma}_{\rm fb}=0$.
Since the total mass of stars at the current MW is $\sim(4\mathchar`-6)\times10^{10}~M_\odot$ \citep{blandh16}, and since the expected total
baryon mass is $M_{\rm b}\sim f_{\rm cb}M_{\rm vir,0}\sim10^{11}~M_\odot$, a significant fraction of gaseous matter should be deposited in the
galactic halo region~\citep{tumlinson17}. The small $\dot{\Sigma}_{\rm fb}$ and $\dot{\Sigma}_{\rm cl}$ are preferred to realize such a situation
under the assumed $\dot{\Sigma}_{\rm b,disk}$ given by the equation~(\ref{eq:sdot_b,disk}).
\par
In this paper we focus on studying the effects of $\eta_{\rm blown}$ and $H_{\rm cr}$. The conversion efficiency $\eta_{\rm blown}$ determines
the relation between the star formation rate, $\dot{\Sigma}_{\rm sf}$, and the disk mass accretion rate, $\dot{\Sigma}_{\rm acc}$. The CR scale
height $H_{\rm cr}$ determines the CR energy density, $e_{\rm cr}$. If the energy density is small, the galactic wind is unable to reach the CGM
region; the wind gas falls back to the disk and the net mass accretion rate increases by the term of $\dot{\Sigma}_{\rm fb}$. Note that the CGM
metal pollution by the galactic wind is required to explain the galactic halo observations and the typical metallicity of the current Galactic disk
of $Z\sim Z_\odot$. $\eta_{\rm blown}$ also determines the mass transfer rate from the disk to the hot gas layer and $H_{\rm cr}$ determines that
from the hot gas layer to the CGM region. The efficient metal pollution of the CGM can result in a large additional mass accretion rate
$\dot{\Sigma}_{\rm cl}$. Thus, these two parameters $\eta_{\rm blown}$ and $H_{\rm cr}$ are important for the gas mass distribution and consequently
for the star formation. We regard the case of $\eta_{\rm blown} =0.1$ and $H_{\rm cr}=10$~kpc as the fiducial model and summarize the other case
in Table~\ref{tb:model}. Note that the values of $H_{\rm cr}$ are chosen to be $(10/7.07)^2\simeq2$ and $(10/14.14)^2\simeq0.5$.
%
\begin{table}[htbp]
\tbl{Model parameters.}{%
\begin{tabular}{c cc c}
\hline
Model & $\eta_{\rm blown}$ &  $H_{\rm cr}$  & category              \\
\hline                                                              
0     & $0.1$              &  $10$~kpc       & fiducial model        \\
1     & $0.2$              &  $10$~kpc       & massive wind          \\
2     & $0.05$             &  $10$~kpc       & weak wind             \\
3     & $0.1$              &  $7.07$~kpc     & large $e_{\rm cr}$    \\
4     & $0.1$              &  $14.14$~kpc    & small $e_{\rm cr}$    \\
\hline                                
\end{tabular}}                        
\label{tb:model}
\end{table}
%
\par
For the numerical calculations, we set the time interval as $\Delta t=0.85$~Myr. Such small $\Delta t$ is required to resolve the lifetime of
massive stars, $\tau_{*,{\rm ms}}\simeq24.2$~Myr. The interval of the radial coordinate for the disk region is set to be $\Delta R=R_{\rm B}/128$.
The initial time is set to $t_0=0.1$~Gyr, and the total baryon mass is given by the $N$-body simulation results. Then, we set $\Sigma(t_0,R)
=$constant with a total mass of $f_{\rm cb}M_{\rm vir}(t_0)\times10^{-2}\simeq7.8\times M_\odot$. The initial mass of the CGM is $M_{\rm cgm}(t_0)
=0.99f_{\rm cb}M_{\rm vir}(t_0)$. The stellar objects and metal masses are initially zero. Note that the results are not affected by these assumed
initial conditions.

\section{Results}
\label{sec:results}
We display the numerical results of our model for parameter the sets summarized by Table~\ref{tb:model}. First, we
present the fiducial model, Model~0. We then discuss the difference between Model~0 and the others.
\par
%
\begin{figure}[htbp]
\begin{center}
\includegraphics[scale=0.65]{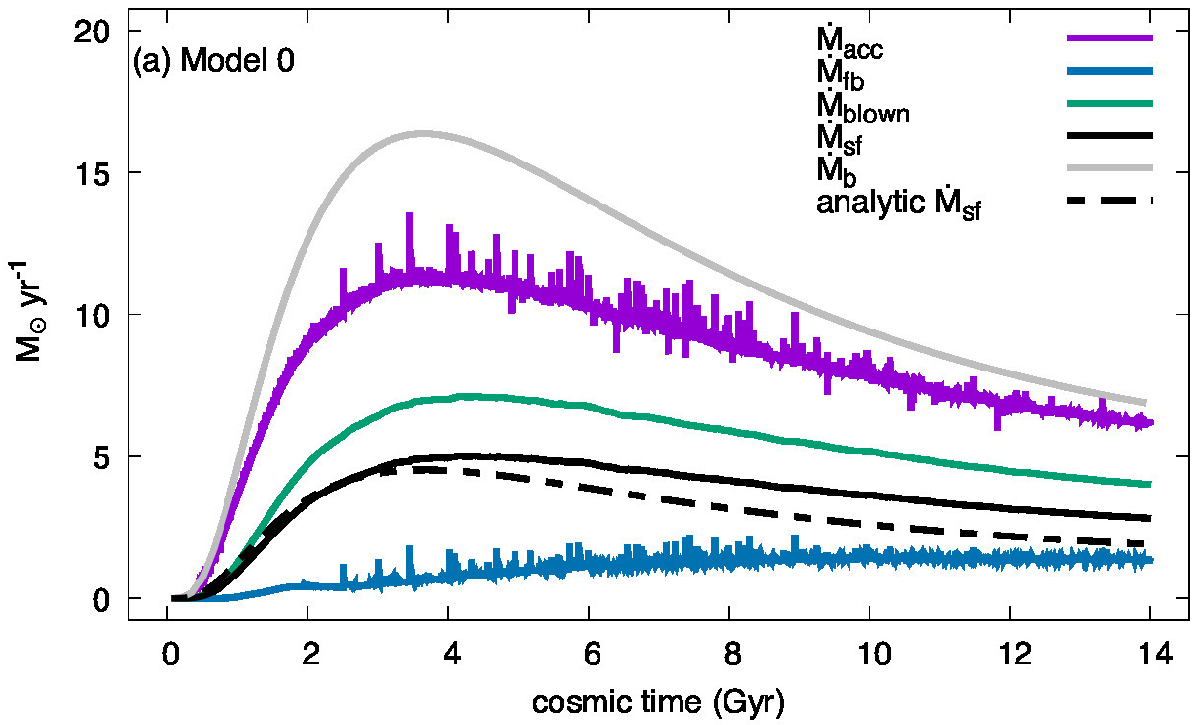}
\includegraphics[scale=0.65]{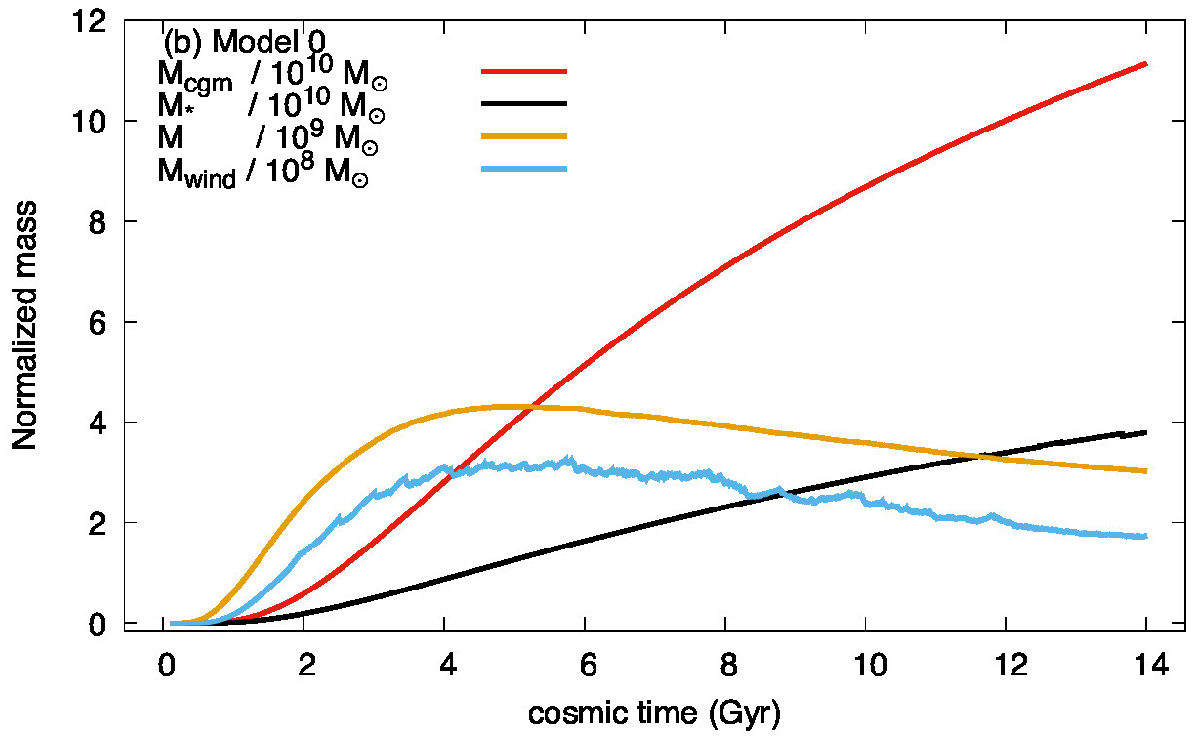}
\end{center}
\caption{The numerical results of Model~0.
(a) The baryon accretion rate $\dot{M}_{\rm b}$ (solid grey line), the estimated star formation rate given
by the equation~(\ref{eq:analytic 2}) with $\dot{\Sigma}_{\rm cl}=\dot{\Sigma}_{\rm fb}=0$ (broken black line),
and the calculated star formation rate $\dot{M}_{\rm sf}$ (solid black line) are also shown in
Figure~\ref{fig:Nbody}a. The solid purple line shows the net accretion rate $\dot{M}_{\rm acc}$. The solid
green line is the outflow rate of the disk $\dot{M}_{\rm blown}$. (b) The results of total mass of the CGM (red),
the stellar mass (black), the gaseous matter at the disk (orange), and the wind (light blue).}
\label{fig:result_model0}
\end{figure}
%
Figure~\ref{fig:result_model0}a shows the results for the total star formation rate, $\dot{M}_{\rm sf}$ (solid black line),
the mass transfer rate given by the falling-back wind particles, $\dot{M}_{\rm fb}$ (purple), and the disk
outflow rate, $\dot{M}_{\rm blown}$ (green).
$\dot{M}_{\rm sf}$ is in good agreement with the estimated star formation rate for which $\dot{\Sigma}_{\rm cl}
=\dot{\Sigma}_{\rm fb}=0$ given by the equation~(\ref{eq:analytic 2}).
$\dot{M}_{\rm fb}$ is small, although its contribution becomes large from $t>8$~Gyr due to a combination of disk mass growth
and CR pressure decrease. Figure~\ref{fig:result_model0}b shows the time evolution of the total mass of the CGM, $M_{\rm cgm}$,
the total mass of low-mass stars, $M_*$,
the total mass of the disk gas, $M$, and the total mass of the wind, $M_{\rm wind}$.
$M_*\simeq4\times10^{10}~M_\odot$ at $t=14$~Gyr is consistent with the current condition of our galaxy \citep{blandh16}.
The time evolution of $M$ is almost the same as $\dot{M}_{\rm sf}$ because of the assumed
$\dot{\Sigma}_{\rm sf}=(\epsilon_{\rm sf}/
\tau_{\rm sf})\Sigma$. The wind mass $M_{\rm wind}$ is flattened from $t\simeq4$~Gyr on, when the growth of virial radius
$r_{\rm vir}(t)$ almost finishes (see, Figure~\ref{fig:Nbody}b). Since we regard the region within $r<2r_{\rm core}(t)=
r_{\rm vir}(t)/10$ as the `wind' region, this feature may have no useful physical meaning.
The mass of the CGM increases monotonically, reflecting the cosmological accretion rate $\dot{M}_{\rm b}$ and
the small mass loss rate of the CGM, $\dot{M}_{\rm cl}$.
\par
%
\begin{figure}[htbp]
\begin{center}
\includegraphics[scale=0.65]{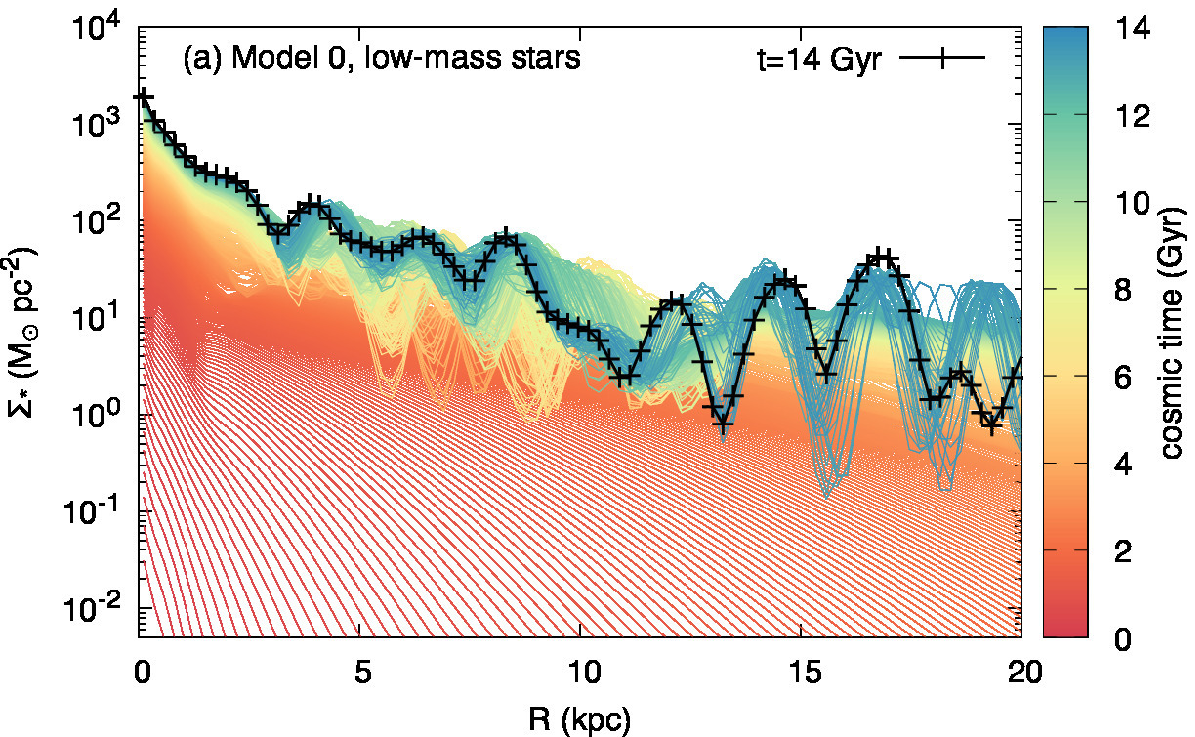}
\includegraphics[scale=0.65]{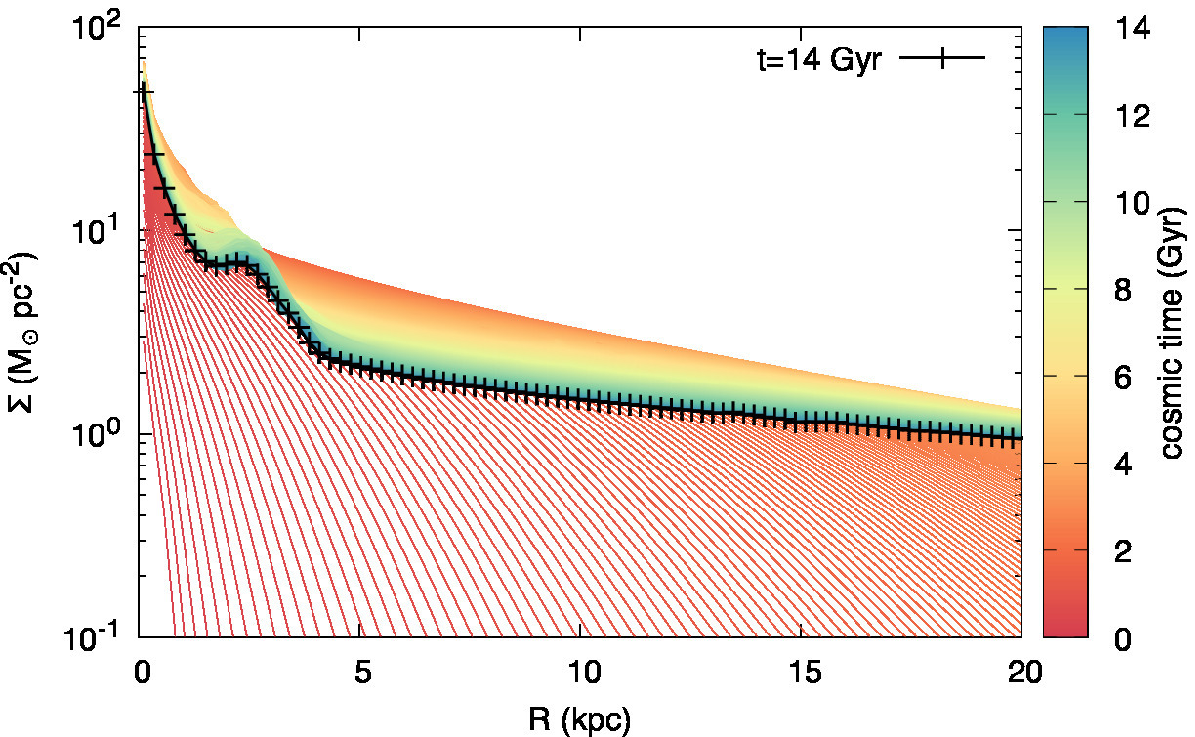}
\includegraphics[scale=0.65]{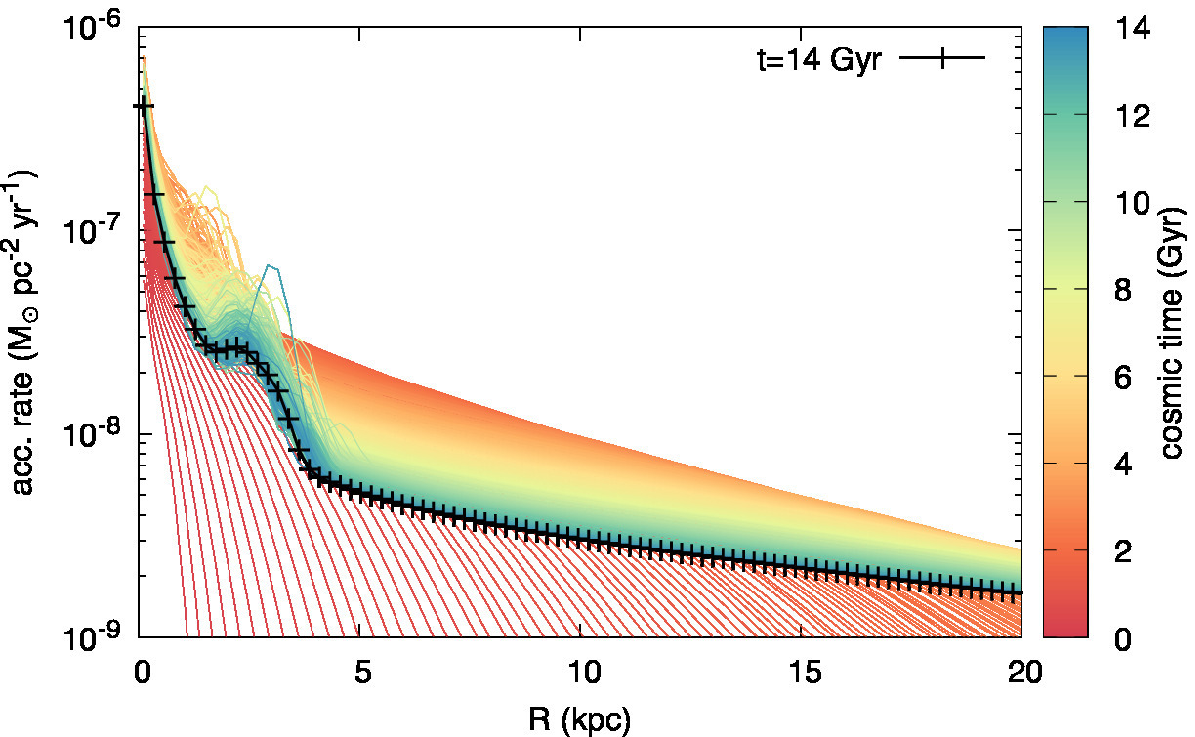}
\includegraphics[scale=0.65]{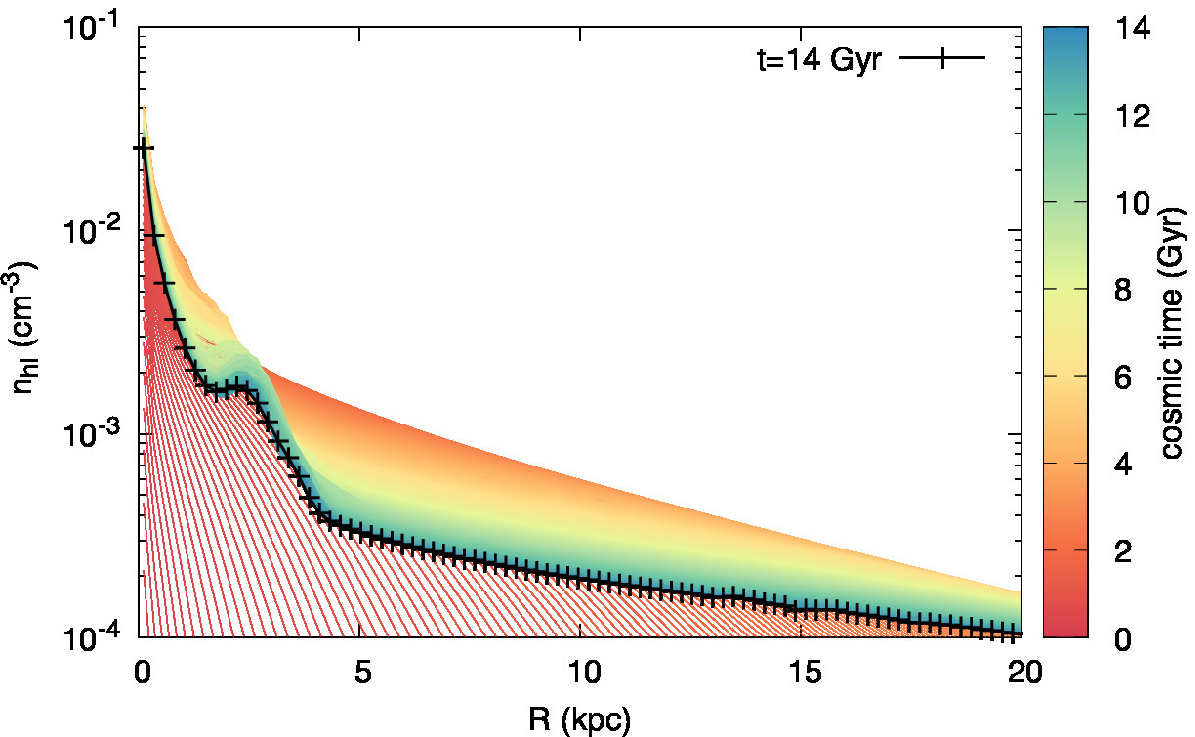}
\end{center}
\caption{The radial profile of the disk and hot gas layer at each time for Model~0.
(a) The surface mass density of the low-mass stars $\Sigma_*$.
(b) The surface mass density of the gas $\Sigma$.
(c) The net accretion rate $\dot{\Sigma}_{\rm acc}$.
(d) The number density of the hot gas layer $n_{\rm hl}\equiv\Sigma_{\rm hl}/(z_{\rm hl}m_{\rm p})$.}
\label{fig:temporal_disk_model0}
\end{figure}
%
Figures~\ref{fig:temporal_disk_model0}a,
\ref{fig:temporal_disk_model0}b, and
\ref{fig:temporal_disk_model0}c show the radial profiles of the surface mass
densities of the low-mass stars, $\Sigma_*(t,R)$, that of the gas, $\Sigma(t,R)$, and the net accretion
rate $\dot{\Sigma}_{\rm acc}(t,R)$ for each time, respectively. The radial profiles of $\Sigma_*$ and $\Sigma$
are nearly converged at $t\sim8$~Gyr, although $\Sigma_*$ is perturbed by the stellar dynamics.
The accretion rate,
$\dot{\Sigma}_{\rm acc}$, shows drastic variations at
$R\lesssim5~{\rm kpc}$ from $t\sim8$~Gyr due to the falling-back of the wind, $\dot{\Sigma}_{\rm fb}$.
Although our treatment of the wind dynamics is still not sufficient (the disk is axisymmetric and
the vertical motion is omitted), this result implies that the inner region of the disk is more
affected by the gas accretion from the galactic halo than the outer region. The variations of $\dot{\Sigma}_{\rm acc}$
correspond to the perturbations of $\Sigma$ which in reality may have a three-dimensional, compact structure in reality
(such as giant molecular clouds). If the case, the local perturbations of the gravitational potential would lead
to the stellar migration phenomena (e.g., \cite{fujimoto23}). It would be an interesting subject to investigate the
birthplace of the solar system, the morphology of galaxies especially for the formation and evolution
of the Galactic bulge region, and so on. Figure~\ref{fig:temporal_disk_model0}d
shows the number density profiles of the hot gas layer, $n_{\rm hl}\equiv\Sigma_{\rm hl}/(z_{\rm hl}m_{\rm p})$,
which can be responsible for the extended soft X-ray emission as we discussed in section~\ref{sec:disk} and
the steady-state solutions of the galactic wind derived by \citet{shimoda22a}.
\par
%
\begin{figure}[htbp]
\begin{center}
\includegraphics[scale=0.75]{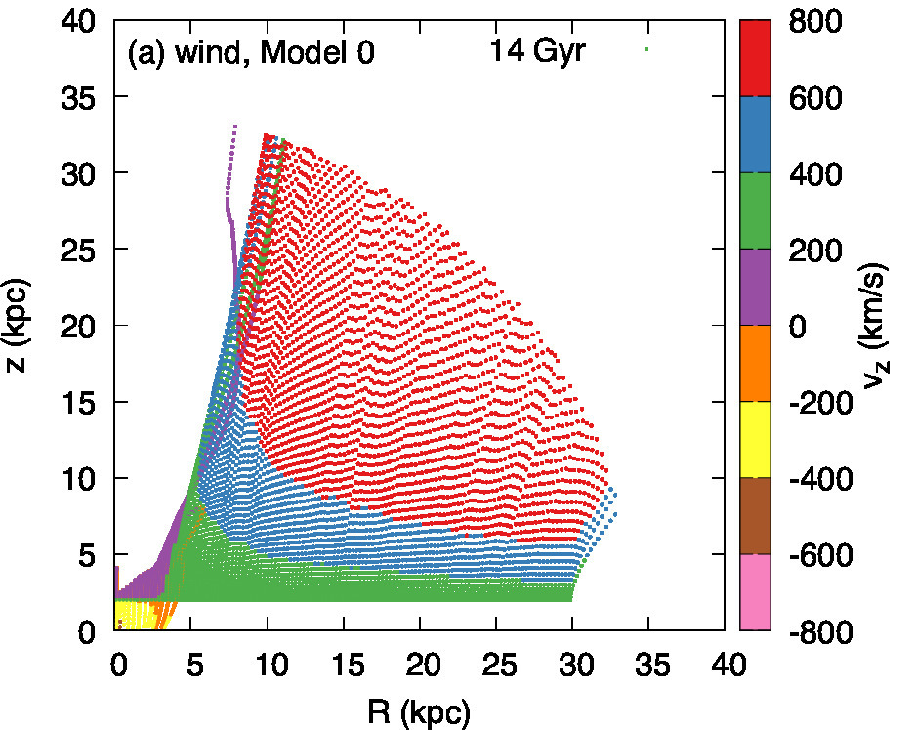}
\includegraphics[scale=0.75]{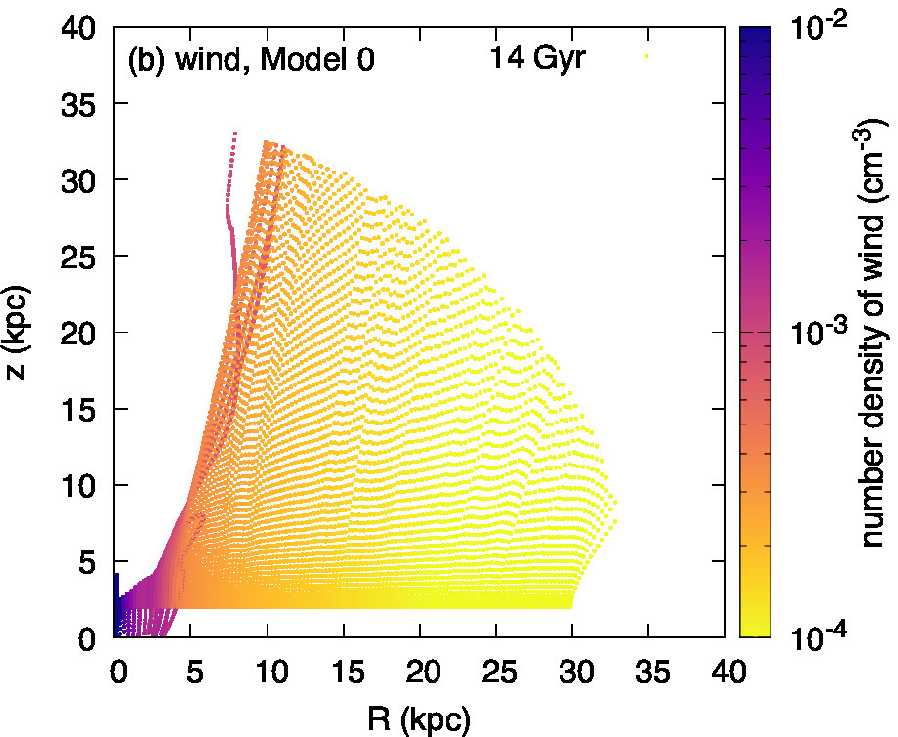}
\end{center}
\caption{The numerical results of the galactic halo region for Model~0.
(a) The wind profile of the vertical velocity component, $v_{{\rm w},z}$, at $t=14$~Gyr.
(b) The wind profile of the number density, $n_{\rm w}\equiv \rho_{\rm w}/m_{\rm p}$, at $t=14$~Gyr.}
\label{fig:wind 14Gyr_model0}
\end{figure}
%
Figures~\ref{fig:wind 14Gyr_model0}a and \ref{fig:wind 14Gyr_model0}b
are the snapshots at $t=14$~Gyr showing the vertical velocity component of the wind, $v_{{\rm w},z}$, and
the number density of the wind, $n_{\rm w}\equiv\rho_{\rm w}/m_{\rm p}$, respectively.  
The wind morphology becomes to the so-called X-shape wind due to the centrifugal force. Such a morphology is reported in
the externalgalaxy of NGC~3079 by \citet{hodges-kluck20}, for example. As shown in Figure~\ref{fig:result_model0}a,
most of the wind mass is transferred to the CGM region ($\dot{M}_{\rm blown}>\dot{M}_{\rm fb}$).
The wind particle that leaves the radius of $R\lesssim5$~kpc eventually falls back to the disk.
This results in the metal pollution of the inner disk region. The metallicity history of the disk
will be discussed later.
\par
%
\begin{figure}[htbp]
\begin{center}
\includegraphics[scale=0.65]{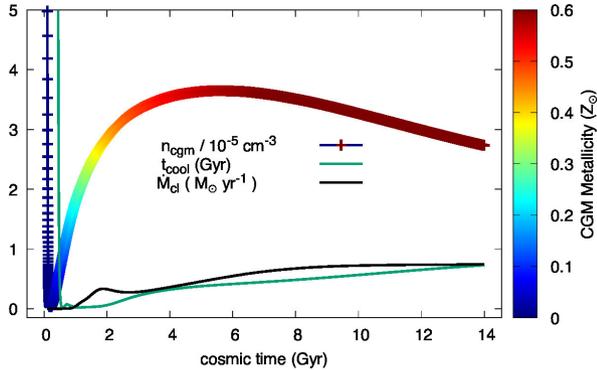}
\end{center}
\caption{The numerical results of the galactic halo region for Model~0.
The time evolution of the CGM number density, $n_{\rm cgm}$, radiative cooling time, $\tau_{\rm cool}$ (green line),
and the mass depletion rate of the CGM, $\dot{M}_{\rm cl}$ (black line). The color bar indicates the metallicity
of the CGM, $M_{Z,{\rm cgm}/M_{\rm cgm}}$.}
\label{fig:cgm 14Gyr_model0}
\end{figure}
%
Figure~\ref{fig:cgm 14Gyr_model0} shows the time evolution of the CGM number density, $n_{\rm cgm}$
(the equation~\ref{eq:n_cgm}), the radiative cooling time, $\tau_{\rm cool}$ (the equation~\ref{eq:t_cool}),
the mass depletion rate, $\dot{M}_{\rm cl}$ (the equation~\ref{eq:mdot_cl}), and the metallicity,
$M_{Z,{\rm cgm}}/M_{\rm cgm}$. At the very early stage $t<1$~Gyr, the very small virial radius, $r_{\rm vir}(t)$,
results in the large $n_{\rm cgm}$. The large $\tau_{\rm cool}$ is due to the small metallicity of the CGM.
At $t>1$~Gyr, $r_{\rm vir}$ becomes $\sim30$~kpc (see, Figure~\ref{fig:Nbody}b), and
then the magnitudes of $n_{\rm cgm}$ and $\tau_{\rm cool}$ converge. The mass growth of the CGM due to the mass transfer
by the wind is balanced with the growth of $\sim r_{\rm vir}(t)^3$ as a result, leading to the almost constant $n_{\rm cgm}(t)
\sim10^{-5}~{\rm cm}^{-3}$. The cooling time becomes $\sim1$~Gyr, which is consistent with the
steady-state solutions of the galactic wind studied by \citet{shimoda22a}. The small mass depletion rate of
CGM, $\dot{M}_{\rm cl}\sim0.5~{\rm M_\odot~yr^{-1}}$, is determined by the assumed efficiency of $\epsilon_{\rm cgm}=0.01$.
The CGM metallicity may be sufficient to reproduce the observed metal absorption lines (see, \cite{tumlinson17} for a review).
\par
%
\begin{figure}[htbp]
\begin{center}
\includegraphics[scale=0.95]{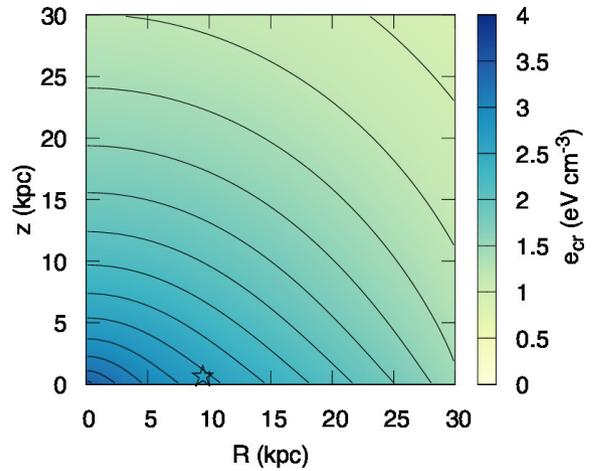}
\end{center}
\caption{The numerical results of the galactic halo region for Model~0.
The profile of the CR energy density, $e_{\rm cr}$, at $t=14$~Gyr.
The black contours show the density
increasing by $0.2~{\rm eV~cm^{-3}}$
from $1~{\rm eV~cm^{-3}}$ to $3.6~{\rm eV~cm^{-3}}$.
The star symbol indicates $R=8.5$~kpc.}
\label{fig:e_cr 14Gyr_model0}
\end{figure}
%
The CR energy density, $e_{\rm cr}$, at $t=14$~Gyr shown in Figure~\ref{fig:e_cr 14Gyr_model0}
is $2\mathchar`-4~{\rm eV~cm^{-3}}$. The radial dependence of $e_{\rm cr}$ reflects the
star formation rate $\dot{\Sigma}_{\rm sf}(R)$ (see, the equation~\ref{eq:cr source}).
Although $e_{\rm cr}$ is measured to be $\simeq 1~{\rm eV~cm^{-3}}$
around the Earth, the actual $e_{\rm cr}$ in the local ISM is still uncertain. To explain the CR ionization rates measured
in the local molecular clouds, the higher energy density of $2\mathchar`-4~{\rm eV~cm^{-3}}$ may be preferred
(e.g., \cite{cummings16}).\footnote{\citet{cummings16} reported the CR energy spectrum in the energy range from $\sim3$~MeV
to $\sim0.3$~GeV by using the {\it Voyger~I} data. These low energy CRs are not expected to penetrate the heliopause.
{\it Voyger~I}, which crossed the solar wind termination shock, measured this energy band for the first time.
They found that the amount of these low-energy CRs is too small to explain the ionization rate of the local molecular clouds.
It is still an open question whether the local clouds consist of a  significant fraction of the CRs with an energy of $<0.3$~MeV,
or whether the CR energy density around the Earth is {\it not} representative of the ISM.} Note that our model is based on
the long-term averaged argument of the star formation with a time scale of $\sim1$~Gyr. The CR energy density measured
around the Earth reflects short-time, local  variations of the ISM such as the local star formation and local diffusion
of the CRs. These possible variations occur on very short-time scales of $\tau_{\rm sf}\sim100~{\rm pc}/C_s
\sim10$~Myr and $(1~{\rm kpc~})^2/{\cal D}_{\rm cr}\sim10~{\rm Myr}({\cal D}_{\rm cr}/3\times10^{28}~{\rm cm^2~s^{-1}})^{-1}$,
respectively.
\par
%
\begin{figure}[htbp]
\begin{center}
\includegraphics[scale=0.60]{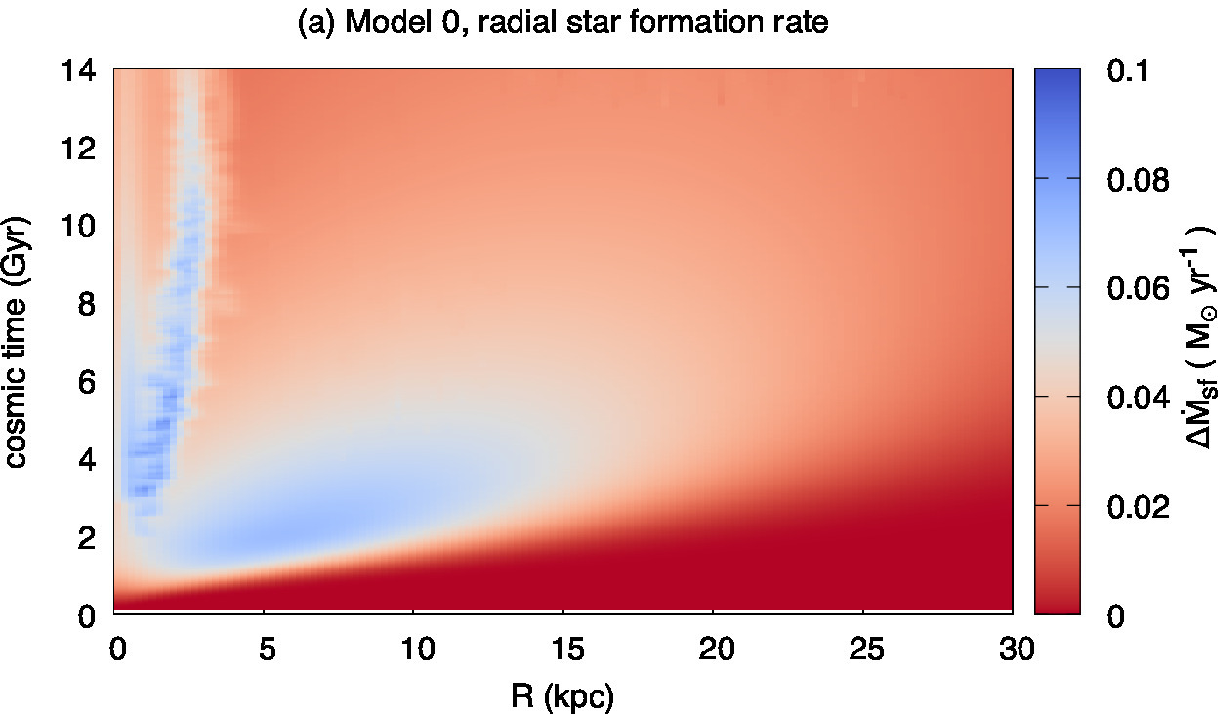}
\includegraphics[scale=0.60]{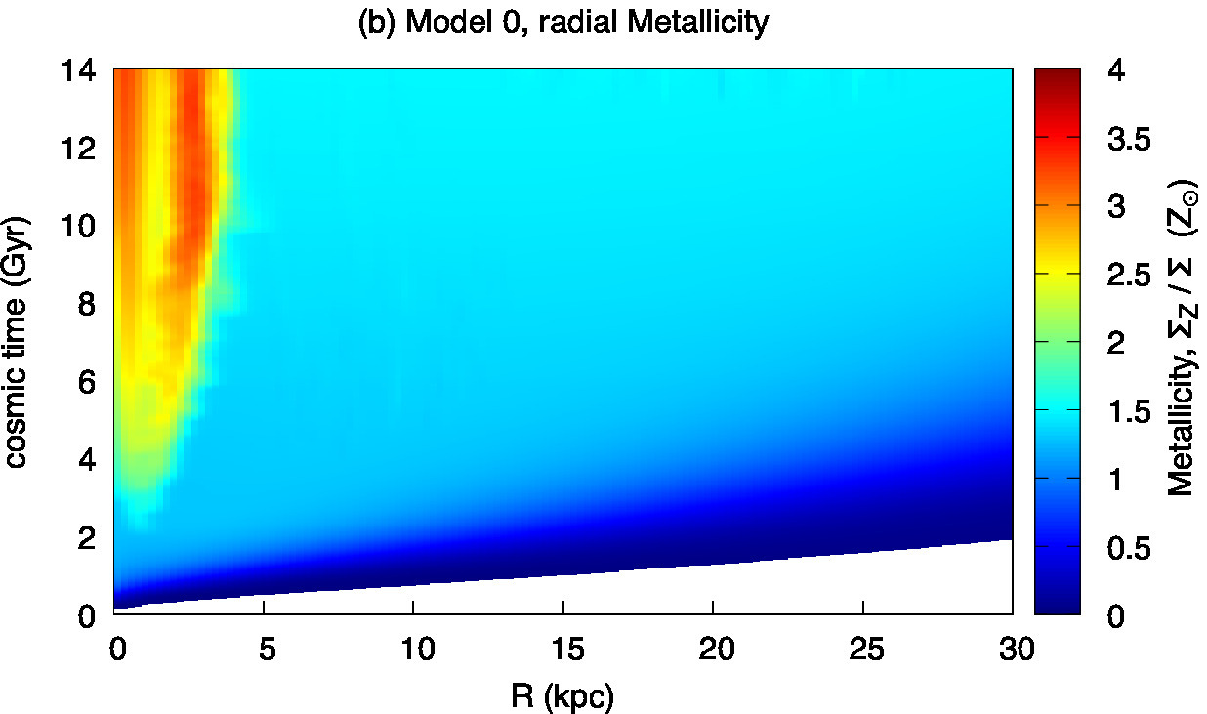}
\includegraphics[scale=0.60]{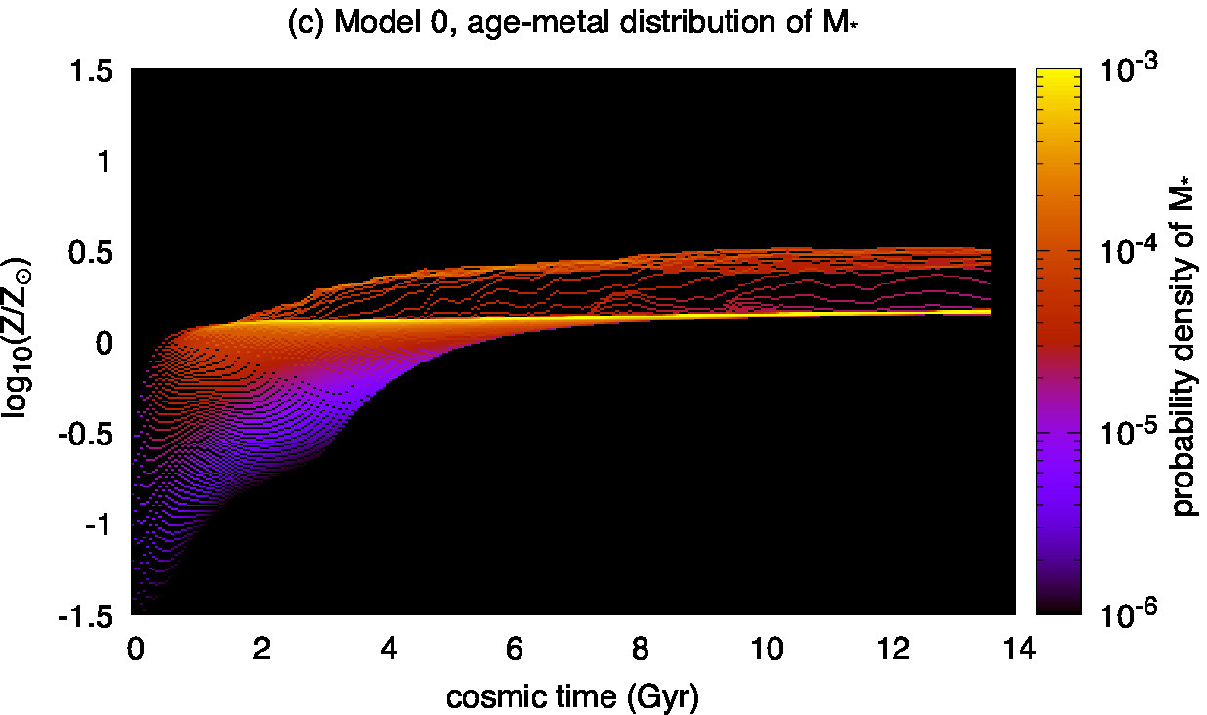}
\includegraphics[scale=0.60]{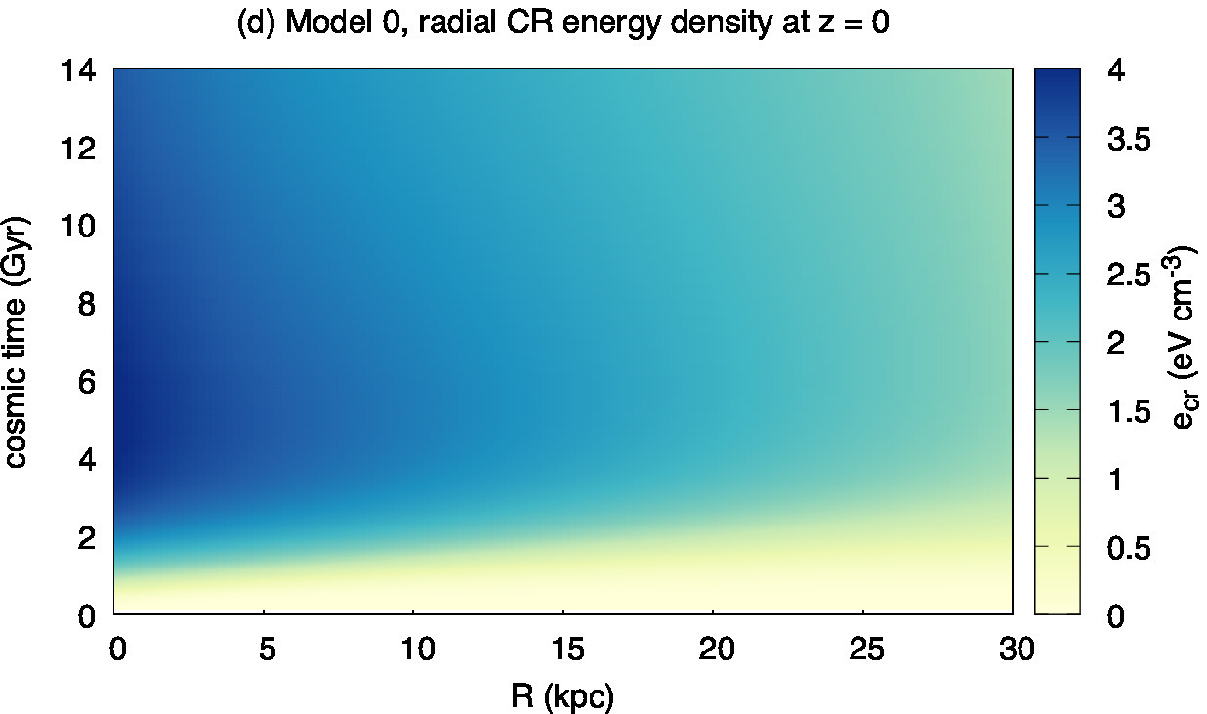}
\end{center}
\caption{(a) The spacetime diagram of the radial star formation rate, $\Delta\dot{M}_{\rm sf}$.
(b) The spacetime diagram of the disk gas metallicity, $\Sigma_Z/\Sigma$ for which
$\Delta\dot{M}_{\rm sf}>10^{-8}~M_\odot~{\rm yr^{-1}}$.
(c) The probability density distribution of the stellar metallicity for the cosmic time,
corresponding to the age-metal distribution of the low-mass stars. (d) The spacetime diagram
of the CR energy density at $z=0$~kpc.}
\label{fig:history_model0}
\end{figure}
%
Figures~\ref{fig:history_model0}a and \ref{fig:history_model0}b show the spacetime diagrams
of the radial star formation rate, $\Delta\dot{M}_{\rm sf}=2\pi \dot{\Sigma}_{\rm sf}R\Delta R$,
and the disk gas metallicity, $\Sigma_Z/\Sigma$, respectively. The star formation history follows
the assumed $\dot{\Sigma}_{\rm b,disk}$ given by the equation~(\ref{eq:sdot_b,disk}). At the inner radius
of $R\lesssim3$~kpc, the mass transfer by the falling-back wind particles leads to higher star formation rates.
The metallicity of the disk gas becomes large at $R\lesssim3~{\rm kpc}$ due to the metal transfer by
the falling-back wind particles. Note that the cosmological accretion gas dose not
contain the metals. The metallicity at $R\gtrsim5$~kpc is almost constant with a value of $\sim Z_\odot$
over the cosmic time because of the large mass transfer rate of wind, $\dot{M}_{\rm w}\sim\dot{M}_{\rm b}$,
and the small mass depletion rate of the CGM, $\dot{M}_{\rm cl}\ll\dot{M}_{\rm b}$.
Most of the created metals goes to and stays in the CGM; $M_{\rm cgm}\sim10^{11}~M_\odot$
(see, Figure~\ref{fig:result_model0}b) and the CGM metallcity is $\sim0.5~Z_\odot$
(see, Figure~\ref{fig:wind 14Gyr_model0}d). This trend of larger metallicity at the inner radius
region is consistent with the current state of the Galactic disk. Figure~\ref{fig:history_model0}c shows
the age-metallicity distribution of the low-mass stars. The older stars tend to have smaller metallcities
and the typical stellar metallicity is $Z_\odot$, which reflects the metallcity of the disk gas. Note that our model
result is based on the long-term averaged argument. The actual stellar metallicity may reflect
local and short time scale physical processes and conditions, which would produce a larger dispersion
than our result.
The mean behavior,--- rapid growth at $t\lesssim2$~Gyr and plateau distribution
with $Z\sim Z_\odot$ at $t\gtrsim2$~Gyr ---, is similar to the observed one by {\it Gaia satellite}  (\cite{xiang22})
The CR energy density, $e_{\rm cr}$, shown in Figure~\ref{fig:history_model0}d also traces
the mass accretion via the star formation rate, but the structures are smoothed out by the diffusion.
\par
%
\begin{figure}[htbp]
\includegraphics[scale=0.65]{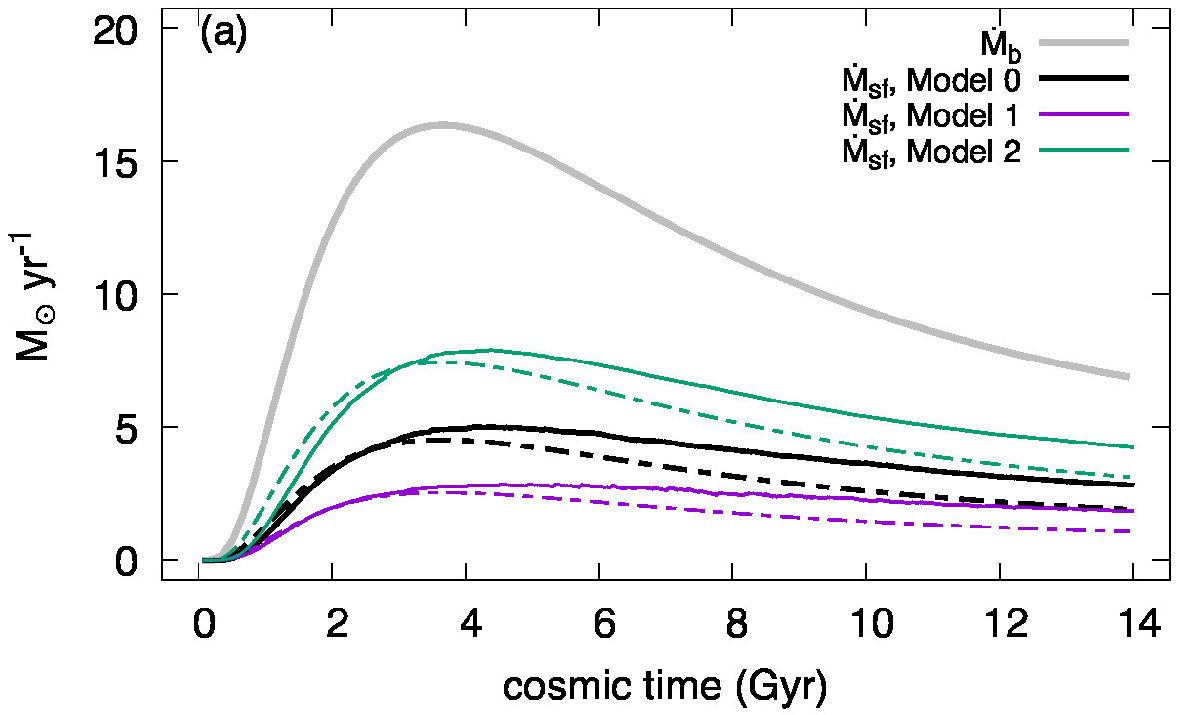}
\includegraphics[scale=0.65]{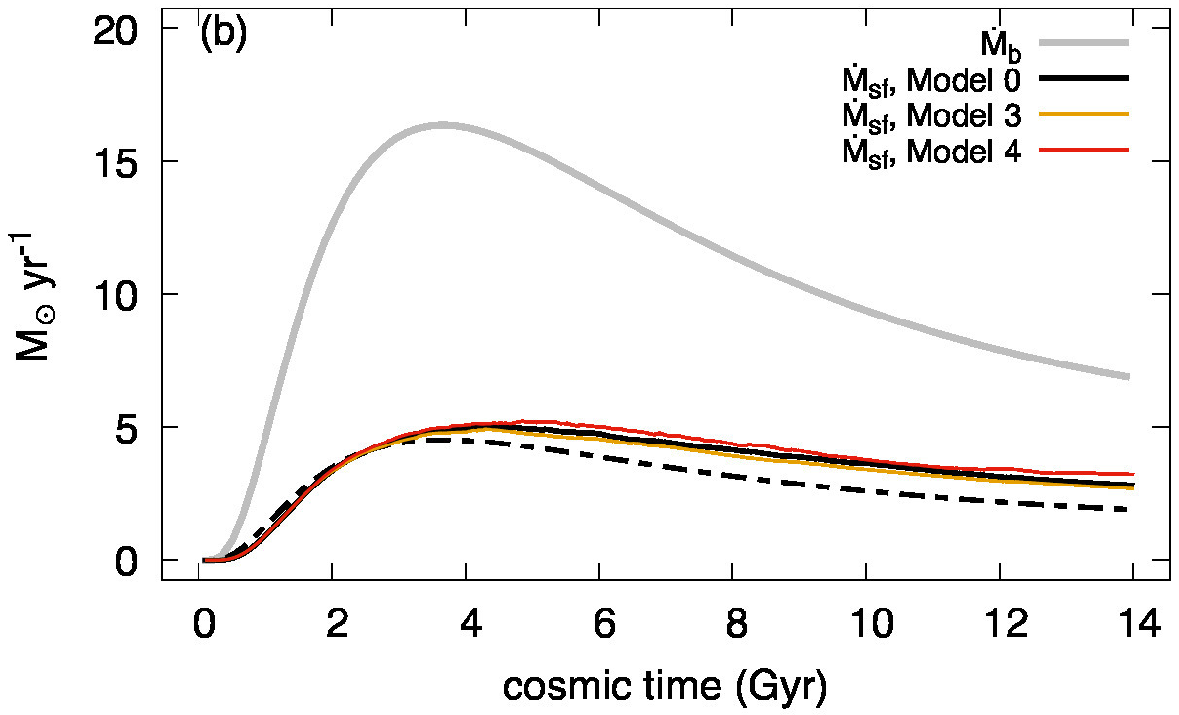}
\caption{The results of star formation rate $\dot{M}_{\rm sf}$ for the models summarized in Table~\ref{tb:model}.
The solid gray line, broken black line, and black lines are the same as Figure~\ref{fig:Nbody}a and \ref{fig:result_model0}a.
(a) The results of the strong wind case (Model~1, purple) and the weak wind case (Model~2, green).
(b) The results of the large $e_{\rm cr}$ case (Model~3, orange) and the small $e_{\rm cr}$ case (Model~4, red).}
\label{fig:sfr_models}
\end{figure}
%
Figure~\ref{fig:sfr_models} shows the results of $\dot{M}_{\rm sf}$ for each model summarized in Table~\ref{tb:model}.
The different conversion efficiency $\eta_{\rm blown}$ (Model 1 and 2) results in the different mass transfer rate  of the outflow via
$\dot{\Sigma}_{\rm blown}=\eta_{\rm w}\dot{\Sigma}_{\rm sn}\propto\eta_{\rm blown}$ and the effects of $H_{\rm cr}$ (Model 3 and 4)
appear via $\dot{\Sigma}_{\rm fb}$. In the case of larger $\eta_{\rm blown}$ (Model~1, the purple lines in Figure~\ref{fig:sfr_models}a),
the gaseous matter is efficiently removed by the wind and therefore the CGM mass becomes larger than in the case of Model~0.
The opposite behavior is obtained in the case of smaller $\eta_{\rm blown}$ (Model~2, the green lines in Figure~\ref{fig:sfr_models}a).
Such simple interpretations can also be found from the fact that the calculated $\dot{M}_{\rm sf}$ of each case is in good agreement with the
estimated star formation rate given by the equation~(\ref{eq:analytic 2}) for which $\dot{\Sigma}_{\rm cl}=\dot{\Sigma}_{\rm fb}=0$.
For Model~1 ($\eta_{\rm blown}=0.05$), the total mass of low-mass stars is $M_*\simeq6\times10^{10}~M_\odot$. This is still consistent
with the observationally estimated total stellar mass in the MW, $4\mathchar`-6\times10^{10}~M_\odot$
(\cite{blandh16}). On the other hand, for Model~2
($\eta_{\rm blown}=0.2$), the mass is $M_*\simeq2\times10^{10}~M_\odot$, which
is not likely. Figure~\ref{fig:sfr_models}b
shows the larger $e_{\rm cr}$ case (Model~3, the orange line) and the smaller $e_{\rm cr}$ case
(Model~4, the red line). Although the total star formation rate of each case may be acceptable, the CR energy density
in the disk becomes $\sim4\mathchar`-8~{\rm eV~cm^{-3}}$ for Model~3, which is a bit too large.
In the case of Model~4, the energy density
is $\sim0.5\mathchar`-2~{\rm eV~cm^{-3}}$, which may still be acceptable. This weak dependence on $H_{\rm cr}$
implies that the Galactic evolution may be stable for variations in the CR energy density due to the uncertain magnetic
field properties (or the diffusion coefficient). Note that the galactic wind solutions derived by \citet{shimoda22a} exist
in the range of $e_{\rm cr}\sim0.1\mathchar`-10~{\rm eV~cm^{-3}}$. Thus, to reproduce the current MW, the parameter
range of $0.05\lesssim\eta_{\rm blown}\lesssim0.1$ and
$10~{\rm kpc}\lesssim H_{\rm cr}\lesssim14~{\rm kpc}$
with $\eta_{\rm cr}=0.1$ is favored.
\par
%
\begin{figure}[htbp]
\includegraphics[scale=0.65]{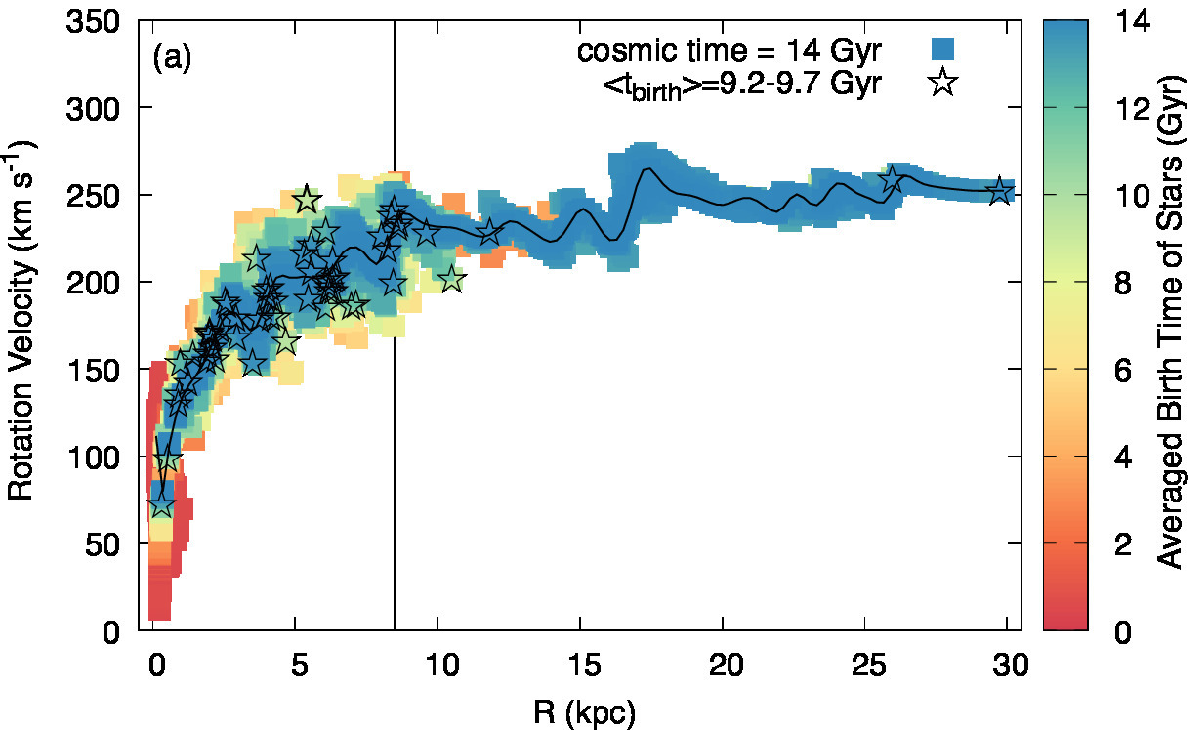}
\includegraphics[scale=0.65]{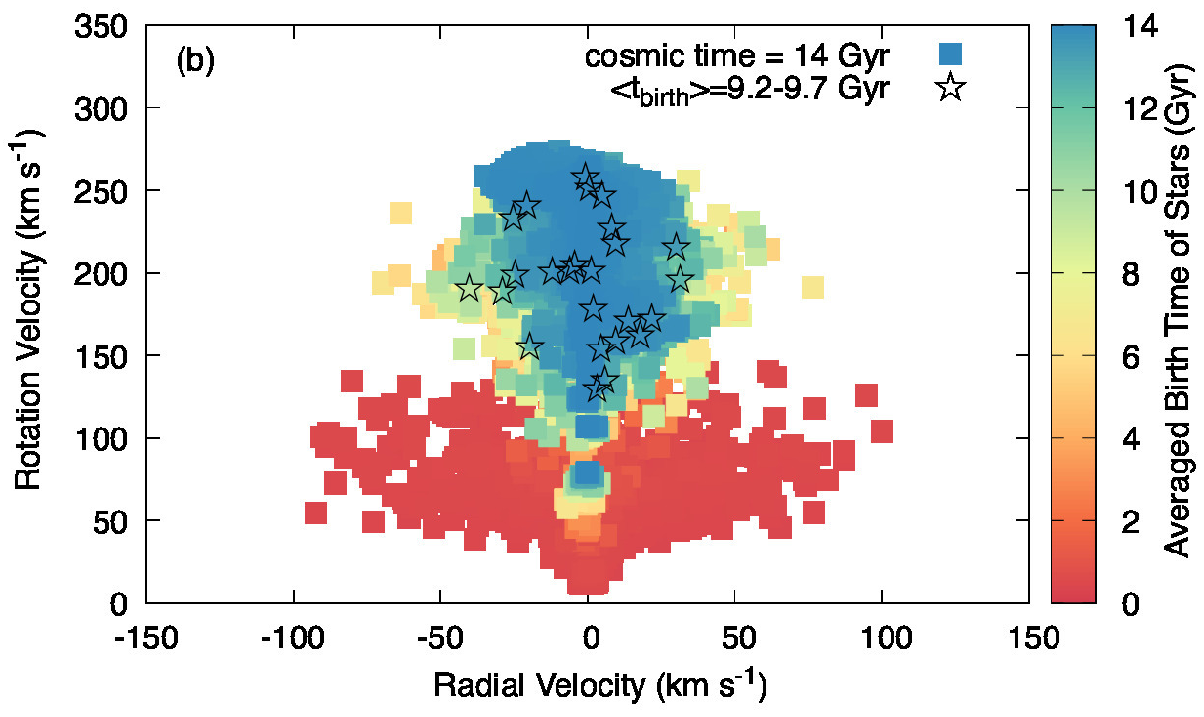}
\includegraphics[scale=0.65]{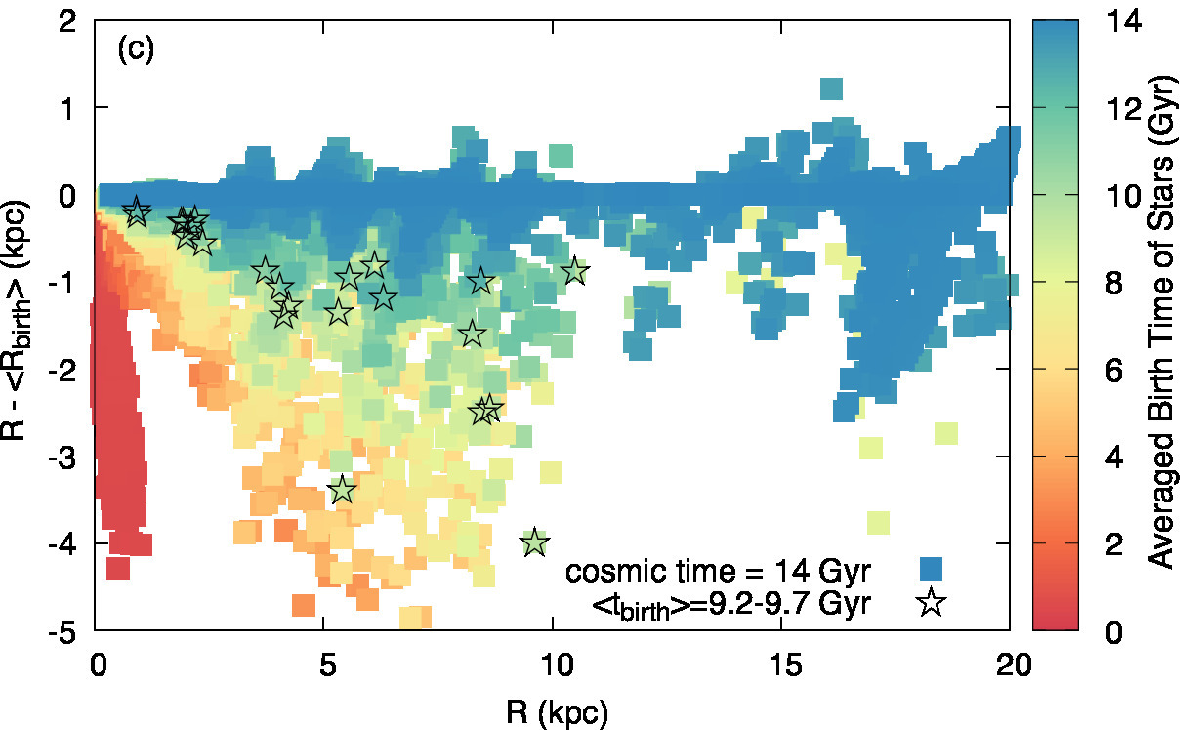}
\caption{(a) The rotation curves at $t=14$~Gyr and $z=0$~kpc of Model~0.
The gray line shows the rotation velocity calculated from the DM potential alone.
The black curve shows the velocities calculated from the sum of the DM and mass components at the disk.
The vertical black line indicates $R=8.5$~kpc.
The square boxes are the stellar parcels and their color shows the averaged birth time, $\langle t\rangle_{\rm birth}$.
The star symbols indicate parcels with $9.2~{\rm Gyr}<\langle t\rangle_{\rm birth}<9.7~{\rm Gyr}$. Note that
the points of the stellar parcels overlap.
(b) The distribution of stellar parcels in the velocity space.
(c) The relation of stellar parcels between the current palce and averaged birth place $\langle R
\rangle_{\rm birth}$.}
\label{fig:rotation_curve}
\end{figure}
%
To test the assumed $\dot{\Sigma}_{\rm b,disk}$ given by the equation~(\ref{eq:sdot_b,disk}), the rotation curve
argument can be useful. Figure~\ref{fig:rotation_curve}a shows the rotation velocity at $t=14$~Gyr and $z=0$~kpc.
The flat rotation with a velocity of $\sim200\mathchar`-250~{\rm km~s^{-1}}$ is well reproduced
(e.g., \cite{mroz19}, for the observed one). The dispersion of rotation curve is due to the collisionless nature of the stars.
For the other models, the curves are similar to each other.
Thus, we regard that the assumptions in $\dot{\Sigma}_{\rm b,disk}$ can be reasonable.
Figure~\ref{fig:rotation_curve}b shows the distribution of stellar parcels in the velocity space
(see, \cite{belokurov18}, for the one observed by the {\it Gaia} satellite).
Note that since our model does not account for perturbations from local objects such as giant molecular
clouds, the result tends to have a smaller dispersion than the realistic one.
The older stars tend to be at $R\lesssim3$~kpc and have a large radial velocity dispersion,
reflecting the rapid growth of the gravitational potential in the early stage. The relatively younger
stars have a smaller radial velocities with a smaller dispersion, corresponding to the Galactic stellar disk.
The growth of gravitational potentail results in a systematic stellar migration which
is indicated by Figure~9c.
This tendency may be an indication of the bulge formation process, which will be studied in our future work.

\section{Implication for Observations}
\label{sec:implication}
We discuss the implications of our model for various observations focusing on
the assumed gas accretion, the origin of the outflowing hot X-ray emitting gas, and so on.
\par
\subsection{The dark accretion flow tested by the X-ray, ultraviolet, infrared and radio observations}
Finding the gas accretion flow onto the disk is important but an unsettled issue in
the MW. In our model, we assume that the most of the cosmological accretion gas with a number
density of $\sim10^{-2}~{\rm cm^{-3}}$ falls directly onto the disk. Although the
actual physical conditions of the cosmological accretion gas are non-trivial, the interactions between
the dense, possibly low-temperature accretion gas and the diffuse gas in the galactic halo
are important issues that may be similar to the cooling flow problem in the clusters of
galaxies (\cite{fabian94,makishima01,peterson03,peterson06}).
If the case, observations at the soft X-ray band may be important to find the accretion flow
in the analogue of the clusters of galaxies.
\par
The high-velocity clouds and intermediate-velocity clouds observed at
the Galactic halo region by the $21~{\rm cm}$ line emission
may be part of this expected accretion gas (\cite{wakker97,putman12,hayakawa22}). 
The estimated mass accretion rate of the observed H\emissiontype{I} components,
$\sim0.1\mathchar`-0.5~M_\odot~{\rm yr^{-1}}$, is too small to be responsible for the
star formation rate of $\sim3~M_\odot~{\rm yr^{-1}}$, although the distance of these clouds are not tightly
constrained (i.e., the total mass is uncertain. See, \cite{putman12}, and references therein).
In our model, the mass accretion from the CGM region occurs
with an assumed efficiency of the angular momentum transfer in condensation processes ($\epsilon_{\rm cgm}=0.01$).
The accretion rate is $\dot{M}_{\rm cl}\sim0.5~M_\odot{\rm yr^{-1}}$, which can be consistent with
the estimated accretion rate of the observed H\emissiontype{I} components. \citet{hayakawa22} argued the origin of
the intermediate-velocity clouds and found that a picture of the external low-metallicity H\emissiontype{I} gas
accretion is favored instead of the Galactic-fountain model (\cite{shapiro76}). This scenario is
consistent with our model results.
\par
Thus, our model predicts
the existence of dark accretion flows (DAFs) with a possible number density of $\sim10^{-2}~{\rm cm^{-3}}$,
corresponding to $\dot{\Sigma}_{\rm b,disk}$ given by the equation~(\ref{eq:sdot_b,disk}).
The DAFs are expected to be responsible for the continuous star formation rate in the disk, and
the galactic rotation curve reflects the accretion dynamics of the DAFs. When the temperature of the DAFs is comparable
to the virial temperature ($\sim10^6$~K), the DAFs are bright in the far ultraviolet
and/or soft X-ray bands, which are actually obscured by the disk gas in the case of observations from the solar system.
Investigating the dynamics of the gaseous matter including the DAFs, wind, H\emissiontype{I} clouds, and so on
would be one of the most important issue.
We will address this issue together with the study of the observational methods in future work.
\par
It is worth emphasizing that an
alternative scenario for the origin of DAFs is that most of the cosmological
gas accretion remains in the Galactic halo or CGM region, --- not accreting directly onto the disk
due to the uncertain angular momentum properties. In this case, the mass accretion responsible for
sustaining the long-term star formation is invoked from the metal-polluted CGM with a larger
$\epsilon_{\rm cgm}$, although most of CGM gas should be virialized to explain the observed massive CGM.
Then, the observational implications are changed as that the accretion gas consists of the metals
and possibly dust grains produced during the gas condensation processes. The metals can produce
the atomic line emissions in the ultraviolet band to the X-ray band so that the gas accretion motions
can be identified by the detailed spectroscopy. The amount of dust grains is also an important predictable
value that can be tested by the observations in the radio and infrared bands.

\subsection{The accretaion history tested by the galactic archaeology}
The relations between the stellar age, metallicity, and phase space distribution
are important clues for investigating a galactic evolution. In the case of the MW,
a sophisticated chemical abundance modeling favors much more episodic gas accretion
onto the disk than that assumed in this paper to explain clear distinguished
two sequences in [Mg/Fe] versus [Fe/H] relation (e.g., \cite{spitoni21,sahlholdt22}).
The interpretation and prediction of the two sequences are not so simple because; (i) the stars can
move from their birth place, (ii) the most of ejected metals should be removed from
the disk by the wind, (iii) a fraction of the wind possibly falls back onto the inner disk,
and (iv) (local) gas or dust migration/dynamics may affect (e.g., \cite{chen23}).
In particular, the formation condition and migration of white dwarf(s) in a close binary system, which results
in the Fe pollution by type~Ia supernovae, may be significantly uncertain. Thus, it would be worth
considering an another footprint for the accretion history.
\par
%
\begin{figure}[htbp]
\includegraphics[scale=0.55]{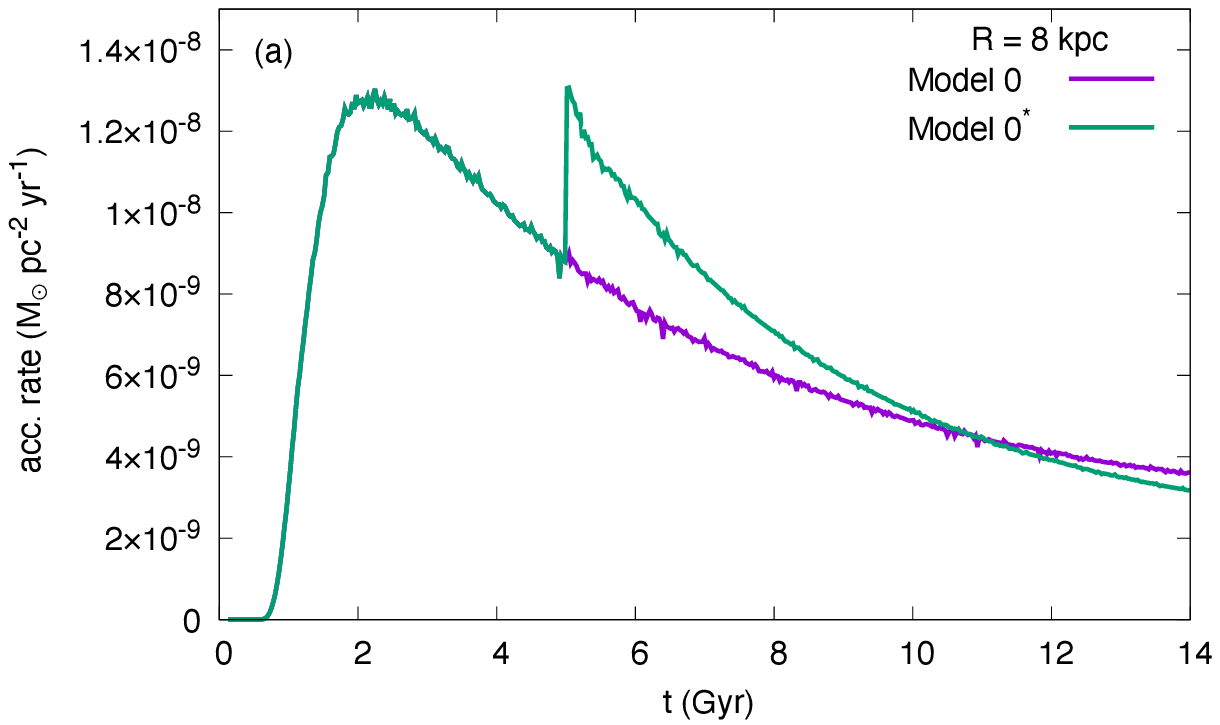}
\includegraphics[scale=0.55]{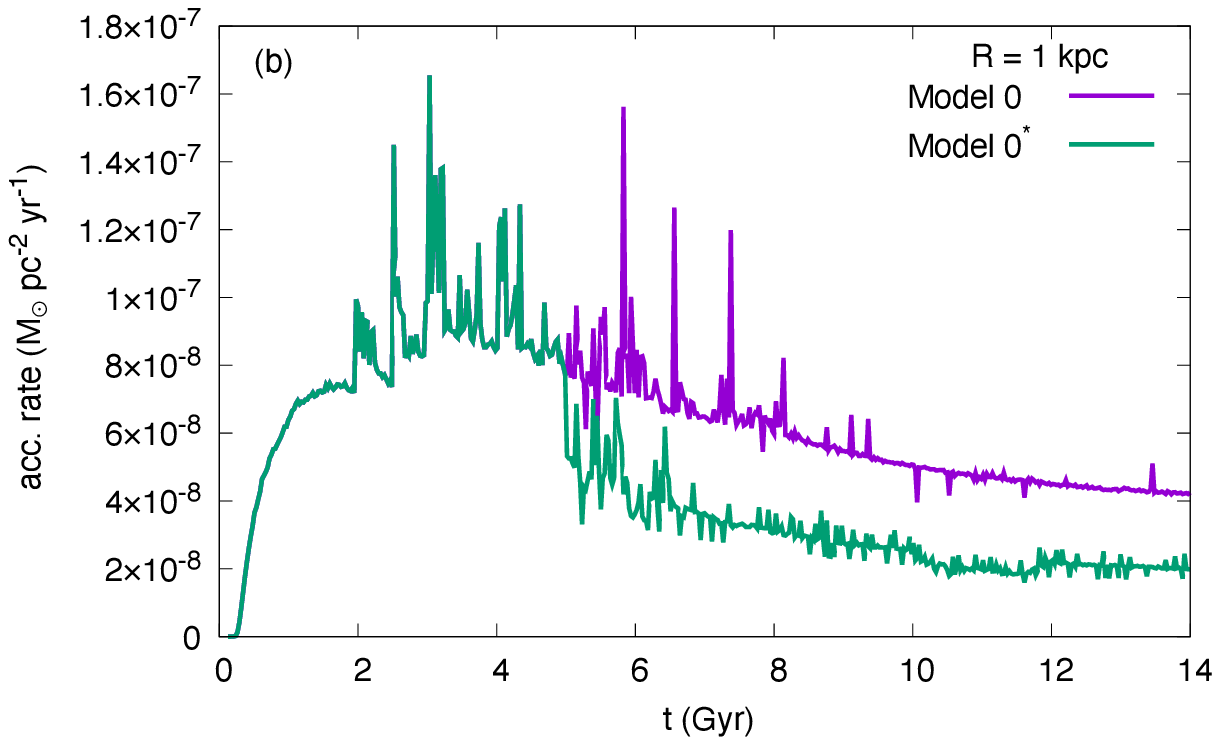}
\includegraphics[scale=0.65]{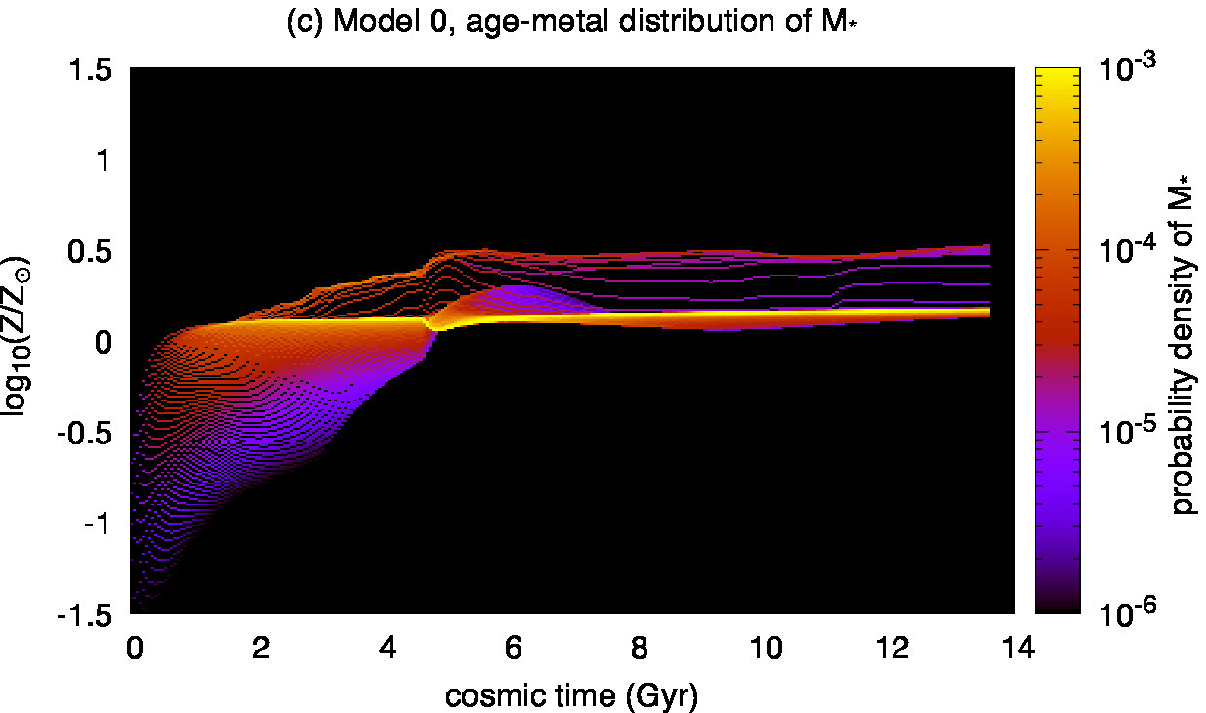}
\caption{(a) The accretion rate onto the disk at $R=8$~kpc. The purple line
indicates Model~0. The green line indicates the case of the episodic accretion.
(b) The rate at $R=1$~kpc.
(c) The age-metal distribution of the low-mass stars in the case of the episodic
accretion. The other parameters are the same as Model~0.}
\label{fig:episodic}
\end{figure}
%
To investigate the accretion history onto the disk, we adopt the episodic gas accretion scenario
by replacing the term of $\exp[-(R-r_{\rm core})^2/2r_{\rm core}{}^2]$ in the equation~(\ref{eq:sdot_b,disk})
to $\exp[-(R-r_{\rm core})^2/2(r_{\rm core}/2){}^2]$ at $t>5$~Gyr, i.e., the accretion gas is more concentrated.
The other parameters are the same as Model~0.
\par
Figures~\ref{fig:episodic}a and \ref{fig:episodic}b are the resultant gas accretion rate
onto the disk at $R=8$~kpc (a) and $R=1$~kpc (b). The concentrated accretion of the primordial gas more dilutes
the disk gas metallicity at $R\sim r_{\rm core}(t=5~{\rm Gyr})\sim10$~kpc, while the inner radius region is less diluted.
Such accretion history forms high-metallicity stars at the inner galaxy which can be seen in the
age-metallicity distribution (Figure~\ref{fig:episodic}c). Future observations providing more complete
stellar samples such as {\it JASMINE} (\cite{gouda21}) will constrain the actual gas accretion history.

\subsection{The origin of outflowing gas tested by X-ray observations}
The gas with the virial temperature for the galactic mass scale, $kT_{\rm vir}\sim0.1~{\rm keV}(M_{\rm vir}/10^{12}~M_\odot)$,
is bright at the X-ray band. The X-ray observations play a key role in studying the outflow
from the galactic system in general.
\par
The origin and formation processes of the diffuse X-ray emission around the Galactic disk are currently
controversial issues as reviewed by \citet{koyama18} for the GDXE.
Note that the outflow is alternatively responsible for removing the metals from the galactic disk.
Thus, a fraction of the disk gas should have a temperature of $T\gtrsim T_{\rm vir}$ to transfer
the metals from the disk to the Galactic halo region.
In our model, we assume that the outflows are driven by consuming a tenth
of the supernova explosion energy ($\eta_{\rm blown}\sim0.1$). The proposed candidates
for the GDXE origin(s) are the assembly of young-intermediate-aged supernova remnants and/or discrete, unresolved
stellar objects. The former scenario is possibly consistent with the numerical simulation results of the local
galactic disk performed by \citet{girichidis18}, although the current computational resources are far from
sufficient to resolve the energy injection processes by the supernovae in general. It is worth pointing out
that the CRs are also able to heat the diffuse gas
via the dissipation of self-excited Alfv{\'e}n waves at a rate comparable to the radiative
cooling rate as discussed by \citet{shimoda22a}. Note that \citet{shimoda22a} assumed one of
the simplest CR heating scenarios (e.g., \cite{volk81,zirakashvili96}), but the effects of the CR heating on
the background plasma remain to be studied (e.g., \cite{zweibel20,yokoyama23}).
Thus, further investigations on the origin of GDXE are necessary for the understanding of the Galactic long-term
evolution and these are the science cases of future X-ray observations such as {\it XRISM} (\cite{xrism20}), and
{\it Athena} (\cite{athena13}).
\par
Related to the origin of the GDXE, the galactic center activity is potentially important
(e.g., \cite{totani06,koyama18}). Note that the galactic center of our galaxy roughly corresponds
to the region around Sgr~A$^*$ within a radius of $\sim300$~pc. This region is not resolved by our model.
The existence of such activities is also favored to explain the bubble-like structures seen in the soft
X-ray sky (e.g., \cite{predehl20}) and possibly smaller bubble structures bright at the $\gamma$-ray band
(the so-called Fermi bubble, \cite{su10}, but see also \cite{crocker22} suggesting the externalgalaxy
scenario). The effects of such activities are not considered in our model and are also important issues
for the net gas accretion rate at the Galactic center, which is related to the formation and evolution
process of the supermassive black hole.
\par
The outflow originated from the disk may explain the metal pollution of the CGM as we mentioned
in this paper. The actual CGM temperature is also important to understand the observed `stable'
CGM (e.g., \cite{miller15,tumlinson17,nakashima18,das19}). If the temperature of CGM was much smaller
than the virial temperature of $\sim10^6$~K, a significant gas accretion from the CGM would happen.
In the case of the MW, the estimated mass and temperature are $\sim4\times10^{10}~M_\odot$
and $\sim3\times10^6$~K, respectively (\cite{miller15}). \cite{das19} reported that a very
hot phase with temperature of $\sim10^7$~K is possibly co-spatial in the CGM. In the case of
external galaxies, lower-ionized ions are also reported (\cite{tumlinson17}). \cite{shimoda22a}
studied the Galactic wind theoretically and showed that the allowed wind solutions range
from $10^4$~K to $10^7$~K depending on the physical conditions (especially, the CR pressure)
at the hot gas layer ($z\sim2$~kpc). Obviously, the angular momentum distribution is also
important however it is more uncertain than the thermal condition. Further theoretical and observational
investigations of the thermal condition, chemical enrichment, and angular momentum distribution of the CGM
are necessary.

\subsection{Implication for the planet formation and cosmic life}
The amounts of metals and CRs in the disk are important for the planet formation.
The protoplanetary disk (PPD) can be quickly dissipated by too efficient angular
momentum transport in the ideal magnetohydrodynamics regime, so the effects
of finite magnetic resistivity are invoked for long-lived PPDs (e.g., \cite{inutsuka12,tsukamoto22b},
for reviews). The magnetic resistivity is, however, uncertain due to large uncertainties in
the CR ionization rate, the amount of the dust, and the dust size distributions (e.g., \cite{tsukamoto22a}).
Further development along the lines of this study
such as the dust formation and detailed CR transport in the disk
would describe the change due to the Galactic evolution for the environment of PPDs in terms of the amount of
dust and CRs over cosmic time as shown in Figures~\ref{fig:history_model0}b and \ref{fig:history_model0}d, and
provide a useful step for investigations of the birthplace of the solar system. We will attempt to extend
our model in future work.
\par
The almost constant star formation rate of $\sim M_\odot~{\rm yr^{-1}}$ corresponds to the almost
constant supernova rates under the assumption of the universal initial mass function and star formation
efficiency, $\epsilon_{\rm sf}$.
In contrast, the size of the gas disk can evolve in time
from the center to outward, reflecting the gas accretion. Thus, the
star and planet systems formed at a very early cosmic time, say $t<2$~Gyr, are more frequently swept
up by the supernova blast waves than those formed at later times. The irradiation of the CRs and high-energy photons
is also strong at the early time. It may be more severe conditions for cosmic life to survive than the current condition.
The frequent shock propagation may also affect the dust grains in the ISM.
At $t\sim4$~Gyr, the size evolution of the gas disk almost converges
(Figure~\ref{fig:temporal_disk_model0}b), and the local star formation rate starts to increase
at $R\lesssim3$~kpc (gradually decreasing at $R\gtrsim3$~kpc, Figure~\ref{fig:history_model0}a).
These are interesting clues to study the habitable region and time of the Galactic disk.

\subsection{Further improvement by observations of externalgalaxies}
The comparison with the observations of externalgalaxies will also provide improvement of our modeling.
Since our model can simultaneously describe the star formation history,  CR energy density, metallicity, and
total baryon mass of galactic halo, the correlations among the galaxy color magnitude
diagram, $\gamma$-ray luminosity, and neutrino luminosity can be studied in a self-consistent manner.
The origin of high-energy cosmic neutrinos discovered by the IceCube Collaboration (\cite{aartsen13a,
aartsen13b}) is considered as a possible evidence for the existence of hidden CR accelerators
(e.g., \cite{murase22}). Our model is also useful for studying such a big enigma in particle
astrophysics.
\par
The difference and relation between the host galaxy and its satellites
would also be interesting subjects. \citet{ruiz-lara20} suggested that repeated encounters of the
Sagittarius dwarf galaxy with the MW result in star formation enhancements in the past of
$\sim5.7$~Gyr, $1.9$~Gyr, and $1.0$~Gyr ago. We will investigate what insights can be obtained from
such multimessenger astronomy methods, including the effects from satellites in future work.
\par
Once we have a reasonable model for Galaxy evolution, we may predict the future of our Galaxy
(see, e.g., \cite{tutukov00}). For example, according to a simple extrapolation we need ~100~Gyr
to make the mean metallicity on the order of 10\%, which might be too large to test observationally.
However, it might not be impossible to study the effect of such a large metallicity by the systematic
observation of the extreme regions in some external galaxies.

\section{Summary}
\label{sec:summary}
We have constructed the model of the Galactic system that describes the long-term evolution of
the star formation, CRs, metallicity, and stellar dynamics over cosmic time. The Galactic gas disk can be modeled
by the accretion disk with the $\alpha$ prescription however the radial mass transfer is
not so important for the surface mass density profiles due to the limited amount of turbulent radial excursion of the gas.
This can be easily understood by $\alpha/C_s{}^2<0.1$ and $10~{\rm kpc}/C_s\sim1~{\rm Gyr}\left(C_s/10~{\rm km~s^{-1}}\right)$
that corresponds to the sound crossing time of the gas at $10^4$~K, which is essentially determined by the Ly$\alpha$ cooling.
When the galactic wind is driven, the long-term star formation can be simply estimated by the equation~(\ref{eq:analytic 1}).
Considering the consistency between the observed diffuse X-ray emission of the current MW \citep{koyama18,nakashima18,predehl20}
and the steady-state wind solutions given by \citet{shimoda22a}, we have found that if about a
tenth of the supernova explosion energy is consumed to form such diffuse, hot X-ray
emitting gas (i.e., $\eta_{\rm blown}\sim0.1$), the star formation rate of a few $M_\odot~{\rm yr^{-1}}$
can be explained. This hypothesis will be investigated by future X-ray missions; {\it XRISM}, {\it Athena}, and so on.
We have found that the acceptable CR energy density of a few ${\rm eV~cm^{-3}}$ in the disk given by
the CR scale height of $H_{\rm cr}\sim10$~kpc can explain the long-term star formation rate.
In our model, a fraction of the wind mass falls back to the disk.
The mass transfer rate of the falling-back wind is a minor component of the net accretion rate
of the total gas mass as a result, while it is responsible for the metallicity increase of the disk gas
in the inner radius region (see, Figure~\ref{fig:history_model0}b).
To test the validity of our scenario in particular the assumed accretion rate of
the disk given by the equation~(\ref{eq:sdot_b,disk}), we have estimated the rotation
curves of the low-mass stars and found that the parameter set of $(\eta_{\rm blown};H_{\rm cr})
=(0.1,10~{\rm kpc})$ can simultaneously explain the star formation rate, metallicity in the disk,
CR energy density in the disk, and the rotation curve.

\begin{ack}
We thank T. Ishiyama, S. Inoue, M. Nobukawa, K. Nobukawa, S. Yamauchi,
T. G. Tsuru, D. Kashino, and S. Cooray, for useful discussions and suggestions.
We are grateful to the anonymous referee, for his/her comments that further
improved the paper.
This work is partly supported by JSPS Grants-in-Aid for Scientific Research Nos. 20J01086 (JS),
18H05436, 18H05437 (SI), and 20H01950 (MN).
\end{ack}

\appendix 


\bibliography{apj_sjsi22}{}
\bibliographystyle{apj.bst}

\end{document}